\documentclass[nohyper,11pt,letterpaper]{JHEP3}
\usepackage[dvips]{graphics}
\usepackage{amsmath}
\usepackage{amssymb}
\usepackage{epsfig}

\def\barray{\begin{array}}
\def\earray{\end{array}}
\def\be{\begin{equation}}
\def\ee{\end{equation}}
\def\ben{\begin{equation} \nonumber}
\def\een{\end{equation}}
\def\ba{\begin{eqnarray*}}
\def\ea{\end{eqnarray*}}
\def\ban{\begin{eqnarray}}
\def\ean{\end{eqnarray}}

\title{
\vskip -50pt
\begin{small}
\hfill DAMTP-2007-74   \\
\hfill PI-COSMO-50    \\
        \vskip  70pt
\end{small}
Magnetogenesis from Cosmic String Loops
}
\author{Diana Battefeld$^1$\footnote{diana.battefeld@helsinki.fi},
Thorsten Battefeld$^1$\footnote{tbattefe@princeton.edu},
Daniel H. Wesley$^1$\footnote{D.H.Wesley@damtp.cam.ac.uk},
Mark Wyman$^2$\footnote{mwyman@perimeterinstitute.ca} \\
$^1$DAMTP, Center for Mathematical Sciences, University of Cambridge, Wilberforce Road, Cambridge,CB3 0WA, UK \\
$^2$Perimeter Institute for Theoretical Physics, 31 Caroline St. N, Waterloo, ON, N2L 2Y5, Canada }

\abstract{
Large-scale coherent magnetic fields are observed in galaxies and clusters, but their ultimate origin remains a mystery. We reconsider the prospects for  primordial magnetogenesis by a cosmic string network. We show that the magnetic flux produced by long strings has been overestimated in the
past, and give improved estimates.
 We also compute the fields created by the loop population,
and find that it gives 
the dominant contribution to the total magnetic field strength on present-day galactic scales.
We present numerical results obtained by
evolving semi-analytic models of string networks (including 
both one-scale and velocity-dependent one-scale models) in a 
 $\Lambda$CDM cosmology,
including the forces and torques on loops from
Hubble redshifting, dynamical friction, and gravitational wave emission.
Our predictions include the magnetic field strength as a function of correlation length, as well as the volume covered by magnetic fields.
We conclude that string networks could account for magnetic fields on galactic scales,
but only if coupled with an efficient dynamo amplification mechanism.
}

\begin{document}

\newpage
\section{Introduction}
Large-scale, coherent magnetic fields with strengths in the $\mu$G range
are observed in galaxies and clusters 
\cite{Grasso:2000wj,Carilli:2001hj,Giovannini:2006kg, clusters}.
Models that explain the presence of these fields fall into two rough categories: ``primordial" mechanisms or ``dynamo" mechanisms.  The primordial option holds that presently observed fields arose from large-scale, homogeneous fields present soon after the Big Bang.  These primordial fields where diluted by cosmic expansion, then slightly amplified by protogalactic collapse to the field strengths presently observed.  The dynamo option assumes that very weak ``seed" magnetic fieds were created by an as-yet-unknown magnetogenesis mechanism in the early universe, which were later greatly amplified by magnetohydrodynamic (MHD) dynamos operating in spiral galaxies.

There are many proposals but few truly compelling mechanisms for primordial magnetogenesis.  On the one hand, the primordial option provides an explanation of presently observed magnetic fields, and the required primordial field strength just evades structure formation and cosmic microwave background (CMB) constraints \cite{Zeldovich65,Grishchuk69,Peebles:1967a,Rees:1972a,Wasserman:1978a,Barrow:1997mj,Durrer:1998ya}.  On the other hand, it is not clear why homogeneous coherent magnetic fields of the correct strength should be present on super-horizon scales.  Inflation is one possible cause, but to first order in perturbation theory, the vector perturbations required to 
create magnetic fields decay with cosmic expansion. Thus any field
generation caused by inflationary perturbations must be a second order effect
\cite{Kobayashi:2007wd}.  This initial conditions issue can be circumvented if fields formed on sub-horizon scales after the Big Bang, with field strength transferred to the larger scales needed to account for galactic magnetic fields \cite{Grasso:2000wj,Giovannini:2006kg,Kronberg:1993vk,Semikoz:2005ks,Widrow:2002ud,Malyshkin:2002a}.  Generally the fields produced in this way are very weak, and must be amplified through the action of a galactic dynamo or ``inverse cascade" arising from turbulent MHD processes
\cite{Grasso:2000wj,Semikoz:2005ks,Brandenburg:1996fc}.
Ultimately all of these mechanisms come up against the correlation length
problem, which itself arises since causal mechanisms
can only operate on sub-horizon scales: except for redshifts close to decoupling, such mechanisms produce fields on comoving lengths that are too small to explain the correlation length of fields observed in galaxies and clusters.
For most mechanisms, this problem can only be solved by invoking large scale field
averaging, inverse cascades, or the super-horizon correlations produced by cosmic inflation. The speculative nature of these proposals
is an indication of how challenging
it is to generate fields with the proper length scales.  In addition, causal mechanisms tend to create fields with a blue power spectrum. If the magnetic fields were created before big bang nucleosynthesis (BBN), then conversion of stochastic magnetic fields into gravitational waves could lead to violations of BBN  bounds \cite{Caprini:2001nb,Kosowsky:2004zh,Caprini:2005ed}.

In this paper we consider  primordial magnetogenesis
by a cosmic string network. 
Such networks are attractive candidates for magnetogenesis,
 since  long strings typically
  stretch across the cosmological horizon, and so can naturally
create coherent effects over large length scales. 
The motion of strings and loops 
through the primordial plasma also produces the
vector-type perturbations required for generating magnetic fields.  Since the magnetic fields are created after BBN, the BBN constraints on stochastic magnetic fields would not apply.
For these reasons and others, cosmic string magnetogenesis has been investigated in the
past by a number of others \cite{Avelino:1995pm,Dimopoulos:1997df,Davis:2005ih}.  
These works share a common approach, whereby  
the string network generates  vorticity in the primordial plasma,
 which is subsequently converted to a magnetic field through the
Harrison-Rees mechanism \cite{Harrison,Rees}. The conversion occurs 
because electrons and protons experience
differing accelerations due to their
Compton scattering against CMB
photons \cite{Rees}. This differential vorticity in the 
plasma creates a current, which then generates magnetic fields.
This process operates so long as the Compton
scattering of CMB photons from free charged particles is efficient, and so 
can create magnetic fields until decoupling.  Conveniently, in this epoch the
cosmological horizon is sufficiently large to produce magnetic
fields on comoving scales which are relevant for the generation of today's galactic and cluster 
magnetic fields. 

 We employ a combination of analytic and numerical techniques 
to estimate the magnetic fields produced by the string network.  The central challenge is to obtain an accurate estimate of the vorticity produced by the network.  We argue that
 the vorticity produced by long, horizon-sized strings has been overestimated in the past, and
here give improved estimates. We also extend our analysis to include the fields generated by the
cosmic string loops produced in string networks. We find 
that the loops turn out to give
the dominant contribution to the total magnetic field produced by the network.
In addition to predicting the overall magnetic field strength, we use
properties of the string and loop networks to compute the spectrum of field
strength as a function of correlation length, as well as the fractional horizon
 volume coverage of regions in which fields are generated.  We obtain these
predictions from a computer code which implements two semi-analytic string network
models, the one-scale model (OSM) \cite{CaldwellAllen1992}
 and the velocity-dependent one-scale model (VOS) \cite{Martins,Tye:2005fn}.
The code also incorporates the dynamics of individual loops, including
the forces and torques on the loops from Hubble expansion, dynamical friction,
and the emission of gravitational waves.  We make minimal assumptions about
MHD processes, and 
 do not invoke any large scale averaging, inverse cascades, or turbulence.

We find that for reasonable choices of string network parameters, 
the network can create sufficiently strong seed fields on galactic
scales, provided the subsequent dynamo amplification is very efficient.  
Cosmic string magnetogenesis has the
satisfying property that it makes a number of other predictions. We find that
adequate magnetic fields can be generated by cosmic strings
with tension $G\mu/c^2 \gtrsim 10^{-8}$,  a value which is still allowed
by cosmological observations \cite{Wyman:2005tu,Seljak:2006bg,Bevis:2007gh,Siemens:2006yp,DePies:2007bm},
 but may be constrained further by
an array of other cosmological observations set to be made within the
next few years \cite{Mack:2007ae,Kuijken:2007ma,Pogosian:2003mz,Seljak:2006hi,Pogosian:2006hg,Hogan:2006we,Siemens:2006vk}.
  Various loop network models give similar
predictions for the magnetic field strength as a function of correlation length,
and this scale-dependence might be observationally testable. Optimistically,
the discovery of cosmic strings with relevant tensions would greatly bolster
the viability of this mechanism. On the other hand,
if galactic dynamos turn out to be less efficient than we assume,
or if no strings  with a high enough tension are found, then our mechanism is ruled out.
 
We organise this paper as follows: in
Section \ref{s:Vorticity} we describe our estimates of the vorticity generated by long string encounters, and discuss the differences between our long string estimates and those in the
literature.  We also give estimates for the vorticity generated by loops, leaving a detailed discussion of the loop dynamics for Appendix \ref{ap1}.
Armed with expressions for the  vorticity generated by individual strings and loops,
 we give a brief description of the network models that describe the long string and loop populations in Section \ref{s:Networks}.   We then use this information to present analytic and numerical estimates of
the magnetic fields generated by the string network in Section \ref{s:Results}.  We give the magnetic field strength as a function of correlation length, describe the volume fraction covered by the loops, and discuss how our results depend on the various string
and loop network parameters. Some details of the code we developed to
 obtain these results is given in Appendix \ref{ap2:code}, and we tabulate our parameter choices in 
Table \ref{table:Parameters}. Our summary and conclusion appear in Section \ref{s:Conclusions}.
Throughout this work, we set $c=1$ but keep other natural constants (such as $G$) explicit.

\section{Generating Vortices}\label{s:Vorticity}

Long strings generate vorticity by dragging
the plasma as they pass through it, and one computes the
drag force by calculating the impulse given to test particles by the passing
string. This problem is considered in detail in refs. \cite{Vachaspati:1991tt,
Vachaspati:1991sy,Hindmarsh:1994re,Dimopoulos:1997df,Davis:2005ih}, 
from which we draw some essential facts, summarised below.
The simplest situation to study is that of an infinitely
long, straight cosmic string with tension $T$, fundamental
mass per unit length $\mu_0$, and effective mass per unit length $\mu$. 
If  extended along the $z$-axis and moving in the $x$-direction with velocity $v_s$, it 
imparts a velocity $v_y$ to a test particle located far from the
$xz$-plane, directed toward the $xz$-plane and given by 
\begin{eqnarray}\label{eq:StringDeflection}
v_y=\frac{2\pi G\lambda}{v_s\gamma_s}+4\pi G\mu_0 v_s\gamma_s\,.\label{vyrel}
\end{eqnarray}
where $\gamma_s=(1-v_s^2)^{-1/2}$, and $\lambda$ is defined as
\be
\lambda = \mu - T.
\ee
The effective mass per unit length, $\mu$, and tension, $T$, are obtained by 
averaging over the small scale fluctuations, or wiggles, on the string.  They are
related to the fundamental ``bare" tension $\mu_0$ by \cite{Vachaspati:1991tt,Vollick:1992sb}
\begin{eqnarray}
\mu T=\mu_0^2\,. \label{muzeroandmu}
\end{eqnarray}
Typically $\mu\approx 1.9\mu_0$ in the radiation era and $\mu\approx 1.5 \mu_0$ in the matter era \cite{VilenkinShellard2000}.
Equation (\ref{eq:StringDeflection}) is valid when we can work in the weak-field limit of GR, when the test particle is far enough from the string that the wiggles can be effectively averaged out, and when the string is  moving sufficiently
rapidly that the test particle does not move substantially closer 
to the string during the encounter.

Only the first term  in (\ref{eq:StringDeflection}), which dominates for
small string drift velocities $v_s$, is important for this work.  
This term is due to the Newtonian gravitational force of 
the string:  in linearised GR the Newtonian potential is sourced by both energy density and pressure, so the string creates the same acceleration
field as would an infinite rod with linear mass density $\lambda$
in purely Newtonian gravity. The second term in (\ref{eq:StringDeflection}) arises from the 
conical deficit angle  $8\pi G\mu_0$ which the string introduces in the surrounding spacetime.  
Simulations indicate that wiggles move very quickly,
 with an RMS velocity of $v_{RMS} >  0.6$, and that the velocity of strings 
 averaged over a correlation length is quite small, $v_s\equiv\bar{v}_{RMS}\approx 0.15$ \cite{Vachaspati:1991tt, Avgoustidis:2005nv}.
So typically the first term dominates, and since  
 $\gamma_s\approx 1$, we are safe in using the non-relativistic limit.
Therefore we are justified in treating the physics of string encounters
with Newtonian gravity.  Unlike the conical-deficit
approach to long strings, this generalises to loops in a simple and convenient way.
It is also a conservative approach, for including  the deficit angle
contribution leads to stronger plasma flows in the string wake,
enhancing the vorticity so produced.

Using the deflection formula (\ref{eq:StringDeflection}) we re-examine the 
calculation of vorticity created by long strings in Section \ref{sec:straight}.  We then apply the same method to estimate the vorticity
generated by string loops in section Section \ref{sec:rotloop}.  We show that producing vorticity by long strings is less attractive than previously thought, but that string loops provide a well-motivated alternative mechanism.

\subsection{Straight Strings \label{sec:straight}}

Long strings generate vorticity by dragging the plasma
behind them as they pass through it,
or equivalently by producing a velocity 
component $v_x$ on test particles.  Treated as a first-order perturbation, 
the $x$-component of the force on a test particle vanishes 
when averaged over the encounter 
with a string.
At second order one includes the motion of the test particle in the 
 $y$-direction during the encounter, which brings the particle slightly
 closer
to the $xz$-plane during the departure of the string than it was during the approach, resulting in a net
force in the $x$-direction.
The final velocity can be computed \cite{Avelino:1995pm} by going 
to the string rest
frame, where the test particle moves with velocity $v_s$.
Since the Newtonian force is conservative, the magnitude of the test particle's velocity is unchanged by the encounter, so 
\begin{eqnarray}
v_s^2=v_y^2+(v_s-v_x)^2,
\end{eqnarray}
and therefore
\begin{eqnarray}
v_x =  \frac{v_y^2}{2v_s} + O(v_y^3).
\end{eqnarray}
To generate magnetic fields, the passing string must generate
vorticity in the plasma.
After the encounter, matter moves toward the string's trajectory
 with velocity $v_y$, but this flow is not rotational.
The plasma's total momentum vanishes after the
encounter, due to the symmetry of the flow with respect
to the $xz$-plane.
At second order, the flow created by the drag velocity $v_x$ 
has net momentum in the direction of the string movement, but is
 not rotational either. 

\begin{figure}[tb]
\centering
  \includegraphics[width=4 in]{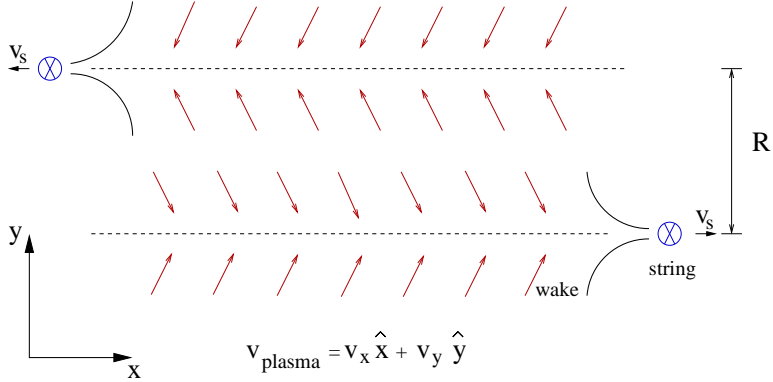}
   \caption{Two straight strings with effective Newtonian mass density $\lambda=\mu-T$ cause wakes in the surrounding plasma via gravitational interaction. After the encounter, the magnitude 
of the dragging component of the plasma flow velocity is approximately $v_y\sim G\lambda/v_s$ and $v_x\sim v_y^2/(v_s)$. The resulting plasma flow carries a rotational component of velocities up to $v_{rot}\sim v_x$ over interstring distances, so that an angular frequency of order $\tilde{\omega}_{pl}\sim \lambda^2G^2/(R_sv_s^3)$ results, see (\ref{omegaplasma2}). \label{fig:strings}}
\end{figure} 

%

A rotational flow can be created by vortices that build up due to turbulence in the string wake, as first proposed by Vachaspati and Vilenkin \cite{Vachaspati:1991tt,Vachaspati:1991sy}  (see also \cite{Vollick:1993ac,Avelino:1995pm}). 
The authors of \cite{Avelino:1995pm} argue that a two-string encounter creates a 
vortex  whose size is comparable to the  
 interstring distance, with rotational velocities comparable to the infall velocity of the plasma $v_{rot}\sim v_y$.   Dimopoulos and Davis \cite{Dimopoulos:1997df,Davis:2005ih} 
later investigated the two-string encounter assuming the same plasma flow.  
The argument they employ runs essentially as follows:
plasma of density $\rho$ in a region of volume $V\sim R_s^2 v_s T$, 
with post-encounter net momentum $\Delta p\sim R_s^3\rho v_x$, leads to
a force on the string of $F\sim \Delta p/T \sim R_s^2\rho v_y^2$.
Applied over a distance $R_s$, the string does work 
$W_s\sim \rho R_s^3 v_y^2$. If we were to assume that the overall flow in the region $V$ is rotational and were to set $E_{rot}\sim \rho R_s^3 v_{rot}^2$ equal to $W_s$, this would imply $v_{rot}\sim v_y$ \footnote{An additional factor of $2$ in \cite{Davis:2005ih} stems from taking $v_{rot}\sim u\equiv2 v_y$, where $u$ is the relative velocity of particles on opposite sides of the string.}.
The keystone of these proposals is the assumption that the
vortex size is comparable to the interstring distance,
and that it carries the majority
of the total kinetic energy imparted to the plasma. As we have shown, such a vortex is not present immediately after the encounter. Indeed, the change in angular momentum of the strings due to the dragging of the plasma is roughly $\Delta J_s\sim R_s \Delta p\sim R_s^4 \rho v_x$. Conservation of angular momentum implies that the plasma may have, at most, a rotational component in the volume $\sim R_s^3$ with angular momentum $J_{plasma}\sim R_s^4 \rho v_{rot} \sim \Delta J_s$. Therefore the net rotational velocity is closer to $v_{rot}\sim v_x$, which is much smaller than $v_y$. 
This is illustrated schematically in Figure \ref{fig:strings}. 

These estimates can be improved as follows.
The wake behind a single string, created at $t_F\lesssim t_{eq}$, has a length $l_w$, width $w_w$ and thickness $d_w$, given by
the scaling relations \cite{Vachaspati:1986,Stebbins:1987cy,VilenkinShellard2000,Pogosian:2004mi}
\begin{eqnarray}
l_w\sim t_F \frac{z_F}{z}\,\,,\,\, w_w\sim v_s t_F \frac{z_F}{z}\,\,,\,\, d_w\sim v_y t_F \left(\frac{z_F}{z}\right)^2 \,,
\end{eqnarray}
which are valid for $z>z_Fv_y/v_s$. Turbulent
eddies arise within the wake shock 
 \cite{Vollick:1993ac}, and could potentially
lead to large
rotational velocities of order $v_y$. However,
the characteristic size associated with
matter chunks due to fragmentation of the wake is
comparable  with the thickness of the wake $d_w$ behind an individual string \cite{Mahonen1993,Avelino:1995zm,VilenkinShellard2000}. We expect turbulent, gravitationally supported vortices at this length scale.
Comparing this thickness to the interstring distance $R_s(t)\sim P^\beta v_s t$ yields
\begin{eqnarray}
\frac{d_w}{R_s}&\sim&\frac{v_y t_F\left(z_F/z\right)^2}{v_sP^\beta t}\label{mentionP}
\sim 2\pi \sqrt{\frac{z_F}{z}}\frac{G\lambda}{v_s^2P^\beta}\,,
\end{eqnarray}
where we use $v_y$ from (\ref{eq:StringDeflection}), $P$ is
the intercommutation probability
\footnote{$P\approx 1$ for cosmic strings, $10^{-1}\leq P \leq 1$
for D-strings and $10^{-3}\leq P \leq 1$ for F-strings
\cite{Polchinski:2004ia}.}, 
 $1/2\leq \beta\leq 1$ \cite{Avgoustidis:2005nv}, and $t_F/t=(a_F/a)^{3/2}\sim (z/z_F)^{3/2}$.
Considering mildly relativistic strings ($v_s\sim 0.1$ as an order of magnitude) 
and the largest possible $G\lambda \sim 10^{-7}$, we are left with  
\begin{eqnarray}\label{eq:WakeOverRs}
\frac{d_w}{R_s}
&\sim&2\pi \sqrt{\frac{z_F}{z}}\frac{10^{-5}}{P^\beta}\,.
\end{eqnarray}
Below we show that the vorticies relevant for magnetogenesis are created toward the end of the radiation era, so 
the redshifting factor is of order unity and vortices due to turbulence in the string wake are much smaller than the interstring distance. 
The seed fields necessary to initiate 
plausible galactic dynamos should be coherent over distances of
at least $\xi_{seed}\sim 5-50\,\text{pc}$ at decoupling, but
$d_{w}\sim 1\,\text{pc}$ for turbulent eddies created around $t_{dec}$.  We
discuss the relevant length scales in more detail in Section \ref{s:Results}.

Since the magnetic field is directly proportional to
the angular velocity of the plasma, our new estimate
greatly reduces the expected 
strength of magnetic fields produced by the long strings. 
Our arguments indicate that the drag velocity $v_x$ is the
relevant velocity for magnetic fields that are coherent over
interstring distances, not the infall velocity $v_y$. This
gives a plasma vorticity
\begin{eqnarray}
\omega_{pl}\sim\frac{v_x}{R_s}\sim \frac{v_y^2}{2R_sv_s}\sim\frac{(2\pi)^2\lambda^2G^2}{2R_sv_s^3} \label{omegaplasma2}\,.
\qquad\text{(long strings, this paper)}
\end{eqnarray}
This is in contrast with \cite{Dimopoulos:1997df,Davis:2005ih}
which obtains the estimate
\begin{eqnarray}
\qquad
{\omega}_{pl}^{\mbox{\tiny\cite{Davis:2005ih}}}\sim\frac{v_y}{R_s}\sim\frac{2\pi\lambda 
G}{v_sR_s}\,.
\qquad\qquad\text{(long strings, previous)}
\label{omegadimopoulos}
\end{eqnarray}
We find that while rotational velocities $\sim v_y$ are possible
within the string wake, by (\ref{eq:WakeOverRs}) their correlation length is much smaller than
the interstring distance $R_s$.

To summarize: vortices due to turbulence in the string wake are confined to
small scales, much smaller than the interstring distance $R_s$.  
Even though these vortices may have large rotational velocities,
their small size makes them far less appealing for magnetogenesis.

\subsection{A Rotating Loop \label{sec:rotloop}}



%
%

\begin{figure}[tb]
  \includegraphics[width=\columnwidth]{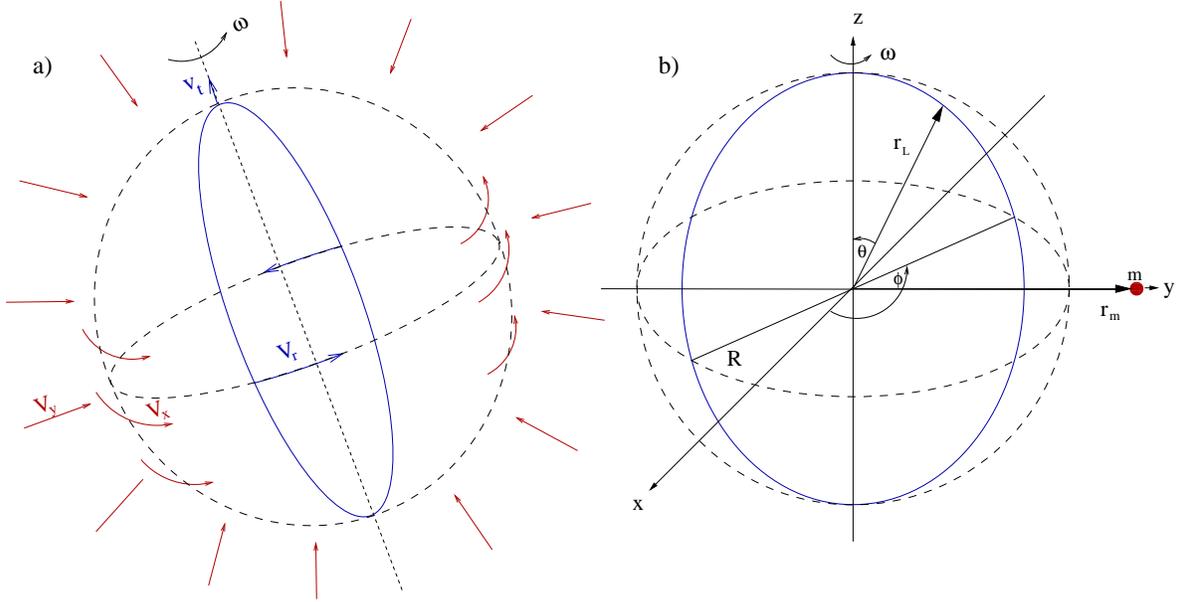}
   \caption{a) A rotating loop with angular velocity $\omega$ and drift velocity $v_t$ attracts the surrounding plasma with a net velocity of order $v_y\sim G\lambda/v_t$, which in turn causes a vortex with rotational velocity of order $v_x\sim v_y^2/v_r$ over the size of the loop. The resulting angular velocity of the plasma is of order $\omega_{pl}\sim\lambda^2 G^2/(\ell v_t^2v_r)$, see eqn.~(\ref{omegapll}). 
b) Shematic of a  rotating loop at ${\bf r}_L(\sigma,t)$ and a test mass at ${\bf r}_m$. Here $\phi=\omega t$, $\theta=\pi\sigma$. \label{fig:loops}}
\end{figure}

We now consider the generation of rotational velocity flows in the plasma
by a loop instead of a long string.  We make the assumption that
 a loop in a typical state of motion has a nonzero angular velocity $\omega$.  The drag force from the loop then transfers angular momentum  to the plasma in a 
straightforward fashion.  The correlation length of the rotational flow, and thus
of the magnetic field, is set by the size of the loop.
In this section we estimate the resulting vorticity.

As in the long string case, our first task is to 
compute the gravitational impulse exerted on a test particle
by a passing rotating loop.
We consider a 
 rigid, circular loop of radius $R$, length $\ell = 2\pi R$, and
linear mass density $\lambda$, with its
angular velocity $\omega$ and translational velocity $v_t$ oriented along the 
$z$-axis. This situation is illustrated in Figure \ref{fig:loops}.
%
%
%
Analytic solutions to the loop equations of motion are known  
\cite{copetur1986}, but in general the loop dynamics are quite complex, and
certainly we do not expect cosmic string loops to act precisely as the
rigid loops we study here.
Nonetheless, we  assume that
the relevant physics operating on the largest loop length scales
is effectively captured by the idealization of a rigid loop, and its
``coarse-grained'' properties such as velocity, angular momentum, mass and size.
The parameters $\ell$, $v_r$ and $v_t$ are all functions of time, 
thanks to the dynamical forces acting on the loop and its
emission of gravitational radiation.
We describe the equations that govern these quantities  in 
Appendix \ref{ap1}, and find that 
$\ell$, $v_r$ and $v_t$ change very little over the timescales
associated with test particle encounters.
While we take all of these dynamical forces
 into account when studying the long-term evolution of
the loop population, for the purpose of
estimating the drag on the plasma, we treat $\ell$, $v_r$ and $v_t$  as constants.

Switching to the loop's rest frame, we consider a particle
at
${\bf r}_m=\tilde{R}\hat{{\bf y}}$. 
At time $t$, we parameterise points ${\bf r}_L$ on the loop by
\begin{eqnarray}
{\bf r}_L=R
\left(\begin{array}{cc}
\sin\pi\sigma\,\cos\omega t\\
\sin\pi\sigma\,\sin\omega t\\
\cos\pi\sigma
\end{array}\right)
\end{eqnarray}
where $\sigma$ ranges over $-1\dots1$. We take the ratio
\be
\mathcal{R}  = \frac{\tilde{R}}{R}
\ee
to be larger than, but close to, unity.  We further define the displacement 
${\bf d}={\bf r}_L-{\bf r}_m$ 
with magnitude
\begin{eqnarray}
d=R\left(1-\frac{2y_m}{R}\sin{\pi\sigma}\sin{\omega t}+\frac{y_m^2}{R^2}\right)^{1/2}\,.
\end{eqnarray}
Recalling that we have defined $x$ as the direction parallel to the
string's motion, we know that the acceleration component $a_x$  vanishes
 to first order if we average over a full rotation. The net acceleration is
in the $y$-direction, toward the loop, and given by 
\begin{eqnarray}
a_y=\pi G\lambda R\int_0^1 \text{d}\tau\int_{-1}^1 \text{d}\sigma\frac{ d_y}{d^3}
 =\mathcal{C}_1\pi \frac{G\lambda}{R},
\end{eqnarray} 
where $\tau=t/T$, with $T$ the loop rotation period
$T=2\pi/\omega$  and
\begin{eqnarray}
\mathcal{C}_1\equiv R^2\int_0^1 \text{d}\tau\int_{-1}^1 \text{d}\sigma\frac{ d_y}{d^3}\,,
\end{eqnarray}
which is of order one. (For example,  one finds $\mathcal{C}_1\approx-1/2$ for $\tilde R=2R$). The net velocity toward the loop after one rotation is then
\begin{eqnarray}
v_y&\approx&\frac{2\pi}{\omega} |a_y|
=|\mathcal{C}_1| 2 \pi^2\frac{G\lambda}{v_r}
\sim  \pi^2\frac{G\lambda}{v_r}\,. \label{vperp1}
\end{eqnarray} 
Thanks to its drift velocity $v_t$, the loop undergoes roughly 
 $4R\omega/(2\pi v_t)=2v_r/(\pi v_t)$ rotations before it  moves away
 from the test particle, so
 the total velocity acquired by the test particle during the encounter is
\begin{eqnarray}
v_y\sim \frac{2\pi G\lambda}{v_t}\,,
\end{eqnarray} 
which is similar to the straight string case in (\ref{eq:StringDeflection}). 
Only the translational velocity $v_t$ enters this expression, since the longer a particle experiences the gravitational attraction toward the loop, the faster they approach each other in the end.

The drifting loop drags the plasma behind it -- just as a straight string does --
but unlike the string encounter, the flow has a rotational component. 
The drag velocity in the $x$-direction is again of order  $v_x\sim v_y^2/v_t$ (see below), resulting in a dynamical friction force  on the loop of $F \sim R^2\rho v_y^2\sim R^2\rho G^2\lambda^2/v_t^2$. 
This has two effects on the loop.
First, it feels a net force due to
dynamical friction of \cite{Chandrasekhar:1943ys,Silk:1984xk}
\begin{eqnarray}
\dot{v}_t=-\frac{v_t}{t_*}\ln \theta_{min}^{-1} \label{accdrag}
\end{eqnarray}
where 
\begin{eqnarray}
\theta_{min}= \frac{2G \lambda \ell}{v_t^2 r_{max}}\label{thetamin}
\end{eqnarray} 
is the minimum scattering angle, $r_{max}=\int v_t \,\text{d}t \simeq 3vt$, and 
\begin{eqnarray}
\label{eq:TStar}
t_*=\frac{v_t^3}{8\pi^2 G^2 R\lambda \rho}.
\end{eqnarray}
Second, because the loop rotates,
the drag force generates
a vortical flow.  
As in the long string case, this is a second-order effect.  The acceleration 
in the $x$-direction due to an infinitesimal
 element of the loop $\text{d}\sigma$ is
\begin{eqnarray}\label{eq:LoopDaccel}
\text{d}a_x&=& \frac{\pi G\lambda R^2}{d^3}\sin{\pi\sigma}\cos{\omega t}
\;\text{d}\sigma
\end{eqnarray}
Substituting the first-order trajectory of the test particle,
given by
\begin{eqnarray}
y_m(t)=\tilde{R}+v_y\left(t-\frac{\pi}{\omega}\right)
\end{eqnarray}
with $y_m(\pi/\omega)=\tilde{R}$, into (\ref{eq:LoopDaccel}),
expanding in terms of $\varepsilon\equiv v_y(t-T/2)/R\ll 1$,
computing the drag velocity by integrating the first order term over a single period
 (the zeroth order contribution vanishes due to symmetry),
and replacing $G\lambda$ in terms of $v_y$
results in
\begin{eqnarray}
v_x &\approx& \frac{v_y^2}{v_r}|\mathcal{C}_2| \sim \frac{v_y^2}{7 \,v_r}, \label{vpara}
\end{eqnarray}
where
\begin{eqnarray}
\nonumber \mathcal{C}_2&\equiv& 12\pi \int_{-1}^{1}\text{d}\,\sigma\int_0^1\,\text{d}\tau\sin{\pi\sigma}\cos{2\pi\tau}\left(\tau-\frac{1}{2}\right)\\
&&\times\frac{\mathcal{R}-\sin{\pi\sigma}\sin{2\pi\tau}}{(1+\mathcal{R}^2-2\mathcal{R}\sin{\pi\sigma}\sin{2\pi\tau})^{5/2}}\,.
\end{eqnarray}
Numerical integration gives
 $\mathcal{C}_2\approx -0.14$ for $\mathcal{R}=\tilde{R}/R=2$, so we
take $\mathcal{C}_2 = -1/7$.
The drift velocity $v_t$ enters (\ref{vpara})
 via $v_y\sim 2\pi G\lambda/v_t$, so as
a result of the drag force, the plasma is stirred up with angular velocity
\begin{eqnarray}
\omega_{pl}\sim \frac{v_x}{\ell}\sim\frac{v_y^2}{7 \ell v_r}\sim\frac{(2\pi)^2\lambda^2 G^2}{7\,\ell v_t^2v_r}\,,\label{omegapll}
\qquad\text{(loops)}
\end{eqnarray}
where we take the vortex size to be given by the loop length $\ell=2\pi R$.
For comparison, the vorticity due to straight string encounter,
derived in 
Section \ref{sec:straight}, is
\begin{eqnarray}
\omega_{pl}\sim\frac{v_x}{R_s}\sim \frac{v_y^2}{2 R_sv_s}\sim\frac{(2\pi)^2 \lambda^2G^2}{2 R_sv_s^3}\,, \label{omegapls}
\qquad\text{(long strings)}
\end{eqnarray}
where $R_s$ denotes the interstring distance
$R_s\approx P^{\beta}v_s t$. 

We show in Appendix
\ref{ap1} that $v_r(t)$ and $v_t(t)$ are
comparable to the average velocity of straight strings,
$v_s$.  Therefore the ratio of vorticities is essentially controlled by
the ratio $\ell/R_s$ of loop length to long string separation.
Since the loop must linger in each region of space long enough to establish
a vortex, vortices are only created
when  $v_r > v_t$.

In this section we focus entirely on the generation of vorticity, since
it is a 
precondition for the Harrison-Rees mechanism to create magnetic flux. 
Since 
these vortices
are real astrophysical objects, they are subject to many physical processes which we have not considered here, and our comparisons
of the length scales and vortex velocities should 
be viewed in this light. For instance, though there is a small vortical component
to the plasma flow in the region between two widely separated cosmic string
wakes, this vast distance will encompass many local over- and under-dense
regions, complicating the physics.
On the other hand, the length scales perturbed by string loops and single string
vortices are somewhat smaller, and should thus be less subject to the 
vagaries of small plasma variations. Furthermore, in each case the strings
are doing more than generating vortical motions and magnetic fields: they are 
accreting matter themselves. This adds to local overdensities, and makes
the regions overswept by strings somewhat more likely later to collapse and
develop structure. We should also have string loops attracted to relatively
overdense areas. We may thus expect regions that have been affected
by string magnetogenesis to be, preferentially, those regions which later
form structure.  Though small, we believe this phenomenon will help to 
increase the effective coverage of string-sourced magnetic fields, since
even if they fail to cover the whole universe, the parts that they do cover
will likely be the parts that will eventually host galaxies.

\section{String Network Models}\label{s:Networks}

To obtain a prediction for magnetogenesis, we must combine our results for
the magnetic fields from a single loop or string,
obtained in 
Section \ref{s:Vorticity}, with a model of the network population.
We use the model to provide the loop length spectrum,
defined by
\begin{equation}\label{eq:LenSpecDef}
N(\ell,t) = \frac{\text{d} N_{loops}}{\text{d} \ln \ell }
\end{equation}
where $\ell$ is the loop length and $N_{loops}$ the number of loops 
per comoving Hubble volume.
To explore the dependence of our results on  
network model assumptions, we use two popular semi-analytic loop
network models: the one-scale model (OSM) and the velocity-dependent
one-scale model (VOS).  Both models obtain equations for the evolution of 
the long string energy density, and then apply energy conservation to 
determine the energy fed into the loop population in the form of 
newly formed loops.

\subsection{The one-scale model (OSM)\label{ss:OSM}}

According to the OSM 
\cite{CaldwellAllen1992,Hindmarsh:1994re,VilenkinShellard2000}
the absolute 
number of loops $N_{loops}$ in a physical volume $V(t)$ obeys the equation
\begin{equation}\label{eq:OSMLoopProdRate}
\frac{\text{d} N_{loops}}{\text{d} L_H(t)} = 
\frac{V(t)}{L_H(t)^4}\frac{C}{\alpha}
\end{equation}
where $C$ is constant during matter- or radiation-dominated epochs, $\alpha$ is the size of newly created loops as a fraction of $L_H(t)$, and
\begin{equation}
L_H(t)= a(t) \int_0^t \frac{\text{d} t'}{a(t')}
\end{equation}
is the particle horizon measured in physical units.  A loop formed at time $t_F$ is taken to have initial length $\ell(t_F)=\alpha L_H(t_F)$, and then subsequently loses energy through the emission of gravitational radiation.  The length at time $t$ of a loop formed at $t_F > t$ is given by
\begin{equation}\label{eq:LoopLenFunTime}
\ell(t,t_F)=f_r \alpha L_H(t_F)-\Gamma_\ell G\mu(t-t_F)
\end{equation}
where $f_r$ represents the energy loss from the redshifting of the loop peculiar velocity immediately after formation, and $\Gamma_\ell$ is a dimensionless parameter controlling the efficiency with which the loop emits gravitational radiation. 

Using the fundamental OSM equations (\ref{eq:OSMLoopProdRate},\ref{eq:LoopLenFunTime}) we now compute $N(\ell,t)$.
To integrate (\ref{eq:OSMLoopProdRate}) we take $V(t)=(a(t)R)^3$, so (\ref{eq:LenSpecDef}) measures the absolute number of loops per (cubical) comoving volume $R^3$.  As we are primarily concerned with
 the radiation dominated epoch, we take $L_H(t) = 2t = 1/H(t)$.
To calculate $N(\ell,t)$ we first rewrite (\ref{eq:LenSpecDef}) as
\begin{equation}\label{eq:OSMtmpSpec}
N(\ell,t) = \frac{\text{d} N_{loops}}{\text{d} L_H(t_F) }
\frac{\text{d} L_H(t_F)}{\text{d} \ln \ell }.
\end{equation}
By (\ref{eq:LoopLenFunTime}), a loop currently of length $\ell$ at time $t$ must have formed at a time $t_F$ when the particle horizon $L_H(t_F)$ was 
\begin{equation}
L_H(t_F) = 2\left(\frac{\ell + \Gamma_\ell G\mu\,ct}{2\alpha f_r + \Gamma_\ell G\mu}\right),
\end{equation}
which yields the second factor in (\ref{eq:OSMtmpSpec}) by differentiation.
To obtain the first factor,
we pick a fiducial time $t_0$ during radiation domination, set 
$a(t_0)=1$, and fix $R=L_H(t_0)$.
This yields the number of loops per logarithmic interval in $\ell$, at time $t$, in a comoving volume equal to one Hubble volume at a time $t_0$, as
\begin{equation}\label{eq:LoopLengthSpectrum}
N(\ell,t) = 
\frac{30}{\alpha}
\left( \alpha f_r + \frac{\Gamma_\ell G\mu}{2}\right)^{3/2}
\frac{L_H(t_0)^{3/2} \ell}{\left(\ell + \frac{\Gamma_\ell G\mu}{2} L_H(t)\right)^{5/2} }
\end{equation}
where we have taken the value $C\sim 30$ during radiation domination, as in  \cite{CaldwellAllen1992}.  
Despite the fact that new loops are being created at $\ell = \alpha L_H$, the spectrum possesses a peak at the characteristic length
\begin{equation}
\ell_{peak}(t) = \frac{\Gamma_\ell G\mu}{2}L_H(t)
\sim 50 \cdot G\mu \cdot ct \label{lpeak}
\end{equation}
taking $\Gamma_\ell \sim 50$ \cite{CaldwellAllen1992,Allen:1994iq,Casper:1995ub}.  The spectrum falls off for $\ell > \ell_{peak}$ since the large loops are created during times of slower loop production, and falls off for $\ell < \ell_{peak}$ since the small loops are near the end of their lives and are evaporating rapidly. Similar expressions can be derived in the matter era.

\subsection{The velocity-dependent one-scale model (VOS)}\label{ss:VOS}

The VOS model of refs. \cite{Martins,Tye:2005fn} characterises the string population
by a length scale $L$, a velocity $v$, and a string number density $n$. 
Commonly, one makes the approximation $n \equiv L^{-2}$, but for our code
we maintain these as distinct parameters for flexibility.
 Like the OSM model, the VOS model has several 
dimensionless parameters, which we call $c_1, c_2,$ and $c_3$.
These are fixed by matching to numerical simulations.
Taking from \cite{Martins} the scaling values 
of $H L$ and $v$, we find $c_1 = 0.21 \;(0.2475), c_2 = 0.18 \;(0.3675)$
in the radiation (matter) eras. We allow free evolution of the network
in the dark energy era using matter era parameters, as this epoch does
not effect magnetogenesis.  The third parameter, $c_3 = 0.28$, fixes the 
scaling string number density. The physical origin of these equations and their
parameter inputs
is discussed extensively in ref. \cite{Martins}; matching between the
equations presented here and those of ref. \cite{Martins} is explained in ref. \cite{Tye:2005fn}.
The VOS model assumes that the length scale 
evolves according to
\begin{eqnarray}
\frac{\text{d} L}{\text{d} t} = H L + c_1 v
\end{eqnarray}
where the loop parameter $c_1 \leq 1$ is  dimensionless.   The velocity
$v$ obeys
\be
\frac{\text{d} v}{\text{d} t} = \frac{-2Hv + c_2/L}{1-v^2},
\ee
and the comoving number density of long strings $N = a^2 n$ is governed by
\be
\frac{\text{d} N}{\text{d} t} = - \frac{c_2 N v}{L} -  \frac{c_3 N^2 L v}{a^2} .
\ee
To translate the output of these equations into what we want -- namely, loop creation rates --
requires a further step. Taking $\rho_\infty = n \mu$, we can just plug the rate of change
of cosmic string energy density into the loop 
creation rate formula from ref. \cite{CaldwellAllen1992},
\be \label{eq:VOSLoopProdRate}
\frac{\text{d} N_{loops}}{\text{d} t} = 
- \frac{V(t)}{\mu \alpha L_H(t)}
\left[ \dot \rho_\infty + 2 \frac{\dot a}{a}\rho_\infty \left(
1 + \langle v^2 \rangle \right) \right],
\ee
where in this formula $L_H$ is again the horizon size, $V$ is the horizon volume, 
and $\alpha$ is, again, the size of the loops after creation as a fraction of $L_H$.

\section{Magnetogenesis and Observational Bounds}\label{s:Results}

We now gather our preliminary results to
determine whether the string network
 can account for the magnetic fields observed today.  In
Section \ref{ss:AfterTdec} we follow the evolution of
the fields from decoupling, through galaxy formation, to the
operation of a dynamo mechanism.  In Section \ref{ss:AnalyticEstimates} we 
give analytic estimates for the resulting galactic fields, followed by
our numerical results in Section \ref{ss:NumericalResults}.
There are great uncertainties affecting almost every stage of 
the evolution of magnetic fields -- especially  the dynamo
amplification efficiency -- but we show that with an efficient dynamo
it is possible to account for presently observed galactic magnetic fields.
We summarise our parameter choices in Table \ref{table:Parameters}, found on 
page \pageref{table:Parameters}.

\subsection{Evolution since matter-radiation equality}\label{ss:AfterTdec}

The vorticity created by the cosmic string network sources magnetic fields
through the Harrison-Rees mechanism
\cite{Harrison,Rees} .
Harrison \cite{Harrison} was the first to consider a model whereby a seed field
is generated by a vortex in the ionized plasma. In Harrison's model, 
a circular current  develops 
when there is a differential rotational flow between ions and the electron-photon gas. Due to
Thompson scattering, non-relativistic electrons are tightly coupled to
the radiation bath before $t_{dec}$. 
Ions are not tightly coupled, so expansion damps their angular momentum
more efficiently.  Because the electrons outpace the ions, a net current
is set up which generates a magnetic field.  Rees \cite{Rees} pointed out that
Harrison's mechanism creates vortices that decay too quickly in the radiation era to produce useful fields, and suggested a modification.
Rees' mechanism relies on
 vortical flows  in the early matter era, when 
vortical motion is less destabilized by expansion.
Since the plasma is still optically thick before recombination,
the angular velocity of the electron fluid is damped through
Compton drag caused by the microwave background;
 meanwhile the relatively heavy ions are less affected.
Again, a current is established, though in a direction opposite
that of Harrison's original proposal.
The strength of a magnetic field produced in this way is 
\cite{Avelino:1995pm}  
\begin{eqnarray}
B =\frac{2m}{e}\omega_{pl}\approx 10^{-4}\omega_{pl}\,,
\end{eqnarray}
where $B$ is in Gauss and $\omega_{pl}$ in $s^{-1}$.
This mechanism can only produce magnetic fields when the universe is ionised and Compton scattering is efficient, which ceases to be true after
decoupling. Though it is also true that the universe
 reionised at low redshift, the 
Harrison-Rees mechanism no longer works in the late universe
since the radiation density is low, rendering Compton scattering ineffective.

Once magnetic fields are produced at a redshift $z_F$, their (proper)
correlation length $\xi$ grows with  the expansion of the universe
as
\begin{eqnarray}
\xi(z) = \frac{1+z_F}{1+z} \xi (z_F),
\end{eqnarray}
and their field strength evolves according to
\begin{eqnarray}
B(z) = \left(\frac{1+z}{1+z_F}\right)^2 B(z_F),
\end{eqnarray}
as a consequence of flux conservation.
Once galaxy formation begins, the evolution of magnetic fields becomes 
far more complicated.  As a 
 protogalactic cloud becomes non-linear and begins to collapse, the 
correlation length decreases but the field strength is amplified. 
While there may be some amplification of the field during collapse 
 \cite{Lesch:1994qb,Kulsrud:1996km}, 
we assume that no dynamo is active at this stage and
the field strength is primarily governed by flux conservation \cite{Turner:1987bw,Davis:1999bt}. The net predynamo amplification factor in a spiral galaxy is approximately \cite{Widrow:2002ud}\footnote{This expression includes: formation of a halo, gain of angular momentum through tidal interactions with neighbouring galaxies, and disc formation, see \cite{Widrow:2002ud} for a review.}.  
\begin{eqnarray}
\frac{B_{i}}{B_{gf}}\approx 8\times 10^3\,,
\end{eqnarray}
where $B_{gf}=B(z_{gf})$.
Once the protogalactic cloud collapses, we assume that
the field is amplified exponentially
by a dynamo mechanism \cite{Widrow:2002ud}, such as 
the $\alpha\omega$-dynamo \footnote{The
$\alpha\omega$-dynamo assumed here applies to galactic magnetic
fields growing in discs only.}.
This dynamo begins operation when stellar winds and explosions
generate interstellar turbulence, which transforms into cyclonic motions
through the Coriolis forces
associated with galactic rotation  
\cite{McIvor,Higdon}. 
The magnetic field surrounding the galaxy has two
modes, a toroidal and a poloidal component. The dynamo converts the
poloidal to toroidal flux by differential rotations of the galactic
disk (the $\omega$-effect) and the toroidal to poloidal through the
cyclonic motions (the
$\alpha$-effect) \cite{Parker}. The combined effects 
can amplify the magnetic field strength by many orders of magnitude
\cite{dynamo,Moffatt,ENParker,Widrow:2002ud}. 

We  parameterise the dynamo
by an efficiency factor $\Gamma_{dy}$, such that $\Gamma_{dy}^{-1}$ is the
field strength
$e$-folding time, and 
the field $B_0$ measured today 
is related to the initial field $B_{i}$ 
by
\begin{eqnarray}
\label{eq:Gammady}
\ln \frac{B_0}{B_{i}} =  \Gamma_{dy}\left(t_f - t_{i}\right) 
\end{eqnarray}
where $t_i\gtrsim t_{gf}$ indicates the onset of dynamo amplification, $t_f\lesssim t_0$ is the time at which the fields reach the observed value, and $t_{gf}$ is the time at which the protogalactic cloud collapses. We take $t_i=t_{gf}$ and  $t_f= t_0$ to arrive at the most optimistic lower bound of the seed field.
The 
resulting amplification is 
weakly sensitive to the choice of $t_{gf}$, and here we take
$t_{gf} = 475$ Myr, which  corresponds to $z_{gf} = 10$.

Although the value of $\Gamma_{dy}$ is crucial to estimate the
necessary seed field, its value is a matter of considerable debate.
In the literature, one finds many
values for $\Gamma_{dy}$ scattered in the range $0.2\,\mbox{Gyr}<\Gamma_{dy}^{-1}<0.8\, \mbox{Gyr}$
\cite{RST,dynamo,Ferriere:2000}.
Recently, some have contended that even larger values, $\Gamma_{dy}^{-1}\gtrsim 1.1-1.4 \,\mbox{Gyr}$, are more likely \cite{Widrow:2002ud}.
Still, taking this uncertainty as an opportunity for optimism, we consider very efficient dynamos with $\Gamma_{dy}^{-1} = 0.2$ Gyr, which amplify by a factor $5.2 \times 10^{28}$
(the amplification factors for various choices of
$\Gamma_{dy}$ and $t_{gf}$ are given in Table \ref{t:DynamoTable}; note that 
in the numerical results that follow, \S \ref{ss:NumericalResults}, $\Gamma_{dy}^{-1}=0.3\,\mbox{Gyr}$ will 
be adequate).  
To obtain the present field of $B_0 = 10^{-6} \, \text{Gauss}$ under these assumptions, 
the field $B_{seed}$ that must be present at decoupling
with $z_{dec} = 1089$ is
\begin{eqnarray}
B_{seed} \approx 2 \times 10^{-35}\, \text{G}\,. \label{boundonbseed}
\end{eqnarray}
This is a very
optimistic lower bound for the seed field at $t_{dec}$. For
 seed fields between this limit and $10^{-20}\,\text{G}$, only the most efficient dynamos might work, though the existence of such dynamos in nature is controversial  \cite{Widrow:2002ud}. 

\begin{table}
\begin{tabular*}{0.90\textwidth}{@{\extracolsep{\fill}}||c|c||r|r|r|r||}
\hline
\multicolumn{2}{||c||}{galaxy formation} & \multicolumn{4}{c||}{$\Gamma_{dy}^{-1}$}  \\
\hline
\multicolumn{1}{||c}{$\quad z_{gf}\quad$} & $t_{gf}$ &  0.2 Gyr & 0.3 Gyr & 0.5 Gyr & 1.0 Gyr \\
\hline\hline
6  & 1 Gyr   & $3.8\times 10^{27}$ & $2.4\times 10^{18}$ & $1.1\times 10^{11}$ & $3.3 \times 10^5$  \\
\hline
10 & $\;$475 Myr $\;$ & $(\star)$  $5.2\times 10^{28}$ &  $1.4\times 10^{19} $ & $3.1\times 10^{11}$ & $5.5 \times 10^5$ \\
\hline
\end{tabular*}
\caption{The amplification factor $B_0/B_{i}$, tabulated with
a variety
  of assumptions regarding the time $t_{gf}$
and redshift $z_{gf}$ of galaxy
formation and the efficiency $\Gamma_{dy}$ of the galactic dynamo. The value we take to get the most optimistic lower bound on $B_{seed}$ in (\ref{boundonbseed}) is marked with a
``$\star$."}
\label{t:DynamoTable}
\end{table}

In addition to constraints on its strength, the seed field must possess a
sufficiently large correlation length.
The correlation length after protogalactic collapse
$\xi_{gf}$  must satisfy
 $\xi_{gf}\geq 100\,\mbox{pc}$ 
for the dynamo to commence \cite{Davis:1999bt}.
Using a simple spherical collapse model for galaxy formation to estimate the comoving correlation length $x_{corr}$ before galaxy collapse leads to \cite{Davis:1999bt}
\begin{eqnarray}
x_{corr}>\eta \,x_{gal}=0.95\,\eta \,(\Omega_m h^2)^{-1/3}M_{12}^{1/3}\,[\mbox{Mpc}]\, \label{xcorr}
\end{eqnarray}
where $M_{12}=M/10^{12}M_{\bigodot}$,
$M$ is the mass of the galaxy, $x_{gal}$ is the comoving length of the
galaxy at formation, 
 and $\eta$ is the fraction of a galaxy over which the magnetic field has to be correlated.
Taking $M_{12} \approx 0.1$ and $\eta \approx 1/150$ (corresponding to  $\xi_{gf}\approx 100\,\mbox{pc}$)  yields
\begin{eqnarray} 
x_{corr}\approx 5.8\,\mbox{kpc}.
\end{eqnarray}
Consequently, the seed fields must have a physical correlation length
$\xi_{seed}$ at decoupling of 
\begin{eqnarray}
\xi_{seed} = \frac{x_{corr}}{1+z_{dec}} > &5.4\,\mbox{pc}\,. \label{xilimit}
\end{eqnarray}
which compares favorably with the particle horizon of $\approx 200$ kpc at
$z \approx 1000$. Larger seed field correlation lengths are even better:
the $5.4$ pc minimum quoted here should cover only about 10\% of the 
protogalactic cloud, which is only marginally adequate. A seed field 
with a longer length scale -- say $50$ pc -- would comfortably 
suffuse the whole protogalactic cloud with a single coherent field.

\subsection{Analytic Estimates}\label{ss:AnalyticEstimates}

\subsubsection{Near decoupling}

Straight string encounters near decoupling produce
vorticity given by $\omega_{pl}$ from (\ref{omegapls}), which by the
Harrison-Rees mechanism creates a seed field of
\begin{eqnarray}
B_{s} \sim 10^{-4} (2\pi)^2\frac{(\mu-T)^2 G^2}{2v_s^3R_s} \label{finalBs} 
\lesssim 1.6 P^{-\beta}\times 10^{-26}\,\mbox{G}\,,\label{finalBsnumber}  
\end{eqnarray}  
where we use $G(\mu-T)\leq 10^{-7}$, $v_s\geq 0.1$ and an interstring distance of $R_s=P^{\beta}v_s t_{dec}$.
These fields have a correlation length at decoupling of
\begin{eqnarray}
\xi_s\sim 12 \, P^{\beta}\,\mbox{kpc}\,.
\end{eqnarray}
The field strength in (\ref{finalBsnumber}) is several orders of
magnitude smaller than the corresponding one in \cite{Davis:2005ih},
since our estimates of the vorticity on large scales generated by a two-string
encounter is much lower than that in \cite{Davis:2005ih}.
Cosmic strings with $P^\beta\sim 1$ and
 the largest possible string tension produce fields that
 are just strong enough to
seed the most efficient dynamos. This improves
somewhat for F and D-strings, which can have a lower
$P^{\beta}$. In either case the coherence length is
larger than the minimal one in (\ref{xilimit}).

The vorticity from rotating loops with $\omega_{pl}$ from (\ref{omegapll})
results in seed fields of 
\begin{eqnarray}
B_\ell&\sim& 10^{-4} \frac{(2\pi)^2}{7} \frac{(\mu-T)^2 G^2}{\ell v_t^2v_r}\,. \label{finalB}
\end{eqnarray}  
Taking $G(\mu-T)\leq 10^{-7}$,  $v_t\geq 0.1$, $v_r\approx 0.4$ and defining a 
new parameter $\tilde\alpha$ such that the loop length $\ell$ at formation is
\begin{eqnarray}\label{eq:TildeAlphaDef}
\ell = \tilde{\alpha}t_{F}=\frac{f_r\alpha}{ H(t_F) }\,,
\end{eqnarray}
at $t_F=t_{dec}$ we obtain 
\begin{eqnarray}
B_\ell \lesssim \frac{2.9 \times 10^{-29}}{\tilde\alpha}\,\text{G}\,.
\end{eqnarray}
We use $v_t\geq 0.1$ since even a large initial velocity of $v_t\sim v_{RMS}$ decreases due to redshifting in the matter era, before it speeds up again due to the rocket effect (see Appendix \ref{ap1:trans}). Because redshifting is absent for the rotational movement, the rotational velocity decreases only very slowly due to the emission of gravitational waves, which is also counterbalanced by loop shrinking to some extent (see Appendix \ref{ap1:shrink}); hence we use use $v_r\approx 0.4$ (Appendix \ref{ap1:rot}). Since $\tilde{\alpha}<P^{\beta}v_s$, we can achieve a larger field strength than for a straight string encounter. 
Consequently, less efficient dynamos work,
but not all galaxies are so lucky as to have had a loop sweeping over them in the past. The coherence length 
\begin{eqnarray}
\xi_l\sim 117 \tilde{\alpha}\, \mbox{kpc}\,.
\end{eqnarray}
is large enough to seed the dynamo for $\tilde{\alpha}\geq 10^{-5}$, 
so that both the largest loops and many of the smaller ones contribute to 
magnetogenesis.  The smaller the loops, the stronger the resulting seed
field.

The analytic estimates suggest similar contributions in magnetic field flux from 
long strings and loops.  In fact, the numerical estimates presented in 
Section \ref{ss:NumericalResults}, which include more details of the loop
dynamics and population characteristics, show that loops produce much
stronger magnetic field fluxes than long strings.  Partly this is because
the  loop length spectrum peaks at lengths much smaller than $\alpha L_H$, which 
effectively lowers the value of $\tilde\alpha$ and greatly increases the 
fields they create.  In addition, redshifting slows the loops, which
then create stronger magnetic fields.

\subsubsection{Fields from before Matter domination}

Extending magnetogenesis into the era before matter domination is a tricky buisness. As one pushes to higher redshift, the horizon becomes ever smaller,
and  sufficient correlation lengths even harder to achieve. Also, having more expansion time for vortices and magnetic fields to dilute pushes the predicted field strengths from these earlier times to even lower and less plausible values. In addition, many dissipative processes are present in this era, causing both $B$ and $\xi$ to decrease further; as a consequence, it is unlikely that any primordial magnetic fields within the horizon survive with reasonable strength and coherence length \cite{Battaner:2000kf}. To complicate matters even more, the full set of MHD equations should be used deep in the radiation era \cite{Dimopoulos:1996nq}, instead of the simple scaling laws. 
We conclude that magnetic fields with the largest fluxes and correlation lengths
 are the ones created around decoupling.

\subsection{Numerical Results}\label{ss:NumericalResults}
The mechanism for magnetogenesis presented here involves too
much physics to be treated efficiently using only analytic methods. But each of the 
relevant physical effects is reasonably well studied, so we include
them using  a computer code. The code is discussed in more detail
in Appendix \ref{ap2:code}. Briefly, the advantages of using the code include:
\begin{itemize}
\item Loops are dynamical objects, and a code can track their evolution over time.
\item The fraction of the universe's volume that is given a seed field is very important
to know, yet very difficult
to estimate analytically. In a code, this can easily be computed.
\item String parameters -- like the bare tension $G\mu_0$, 
and $\alpha$, the loop size -- enter the equations in a variety of places, making it
hard to guess how magnetogenesis strength, length scales, and volume 
coverage will be affected by each. Our code allows us simply to try a variety 
of values and compare the outcomes to develop some intuition about their 
effects.
\item The relative sizes of loop and straight string contributions to magnetogenesis
are easy to calculate and compare on equal footing, without making the possibly 
prejudicial approximations necessary for analytical estimation.
\end{itemize}

\noindent The model parameters we wish to vary are:
\begin{enumerate}
\item $G\mu_0$,
the string's bare tension; 
\item our model for string network evolution;
\item whether or not string loops are allowed to undergo dynamics; 
\item the initial velocity at which a loop moves after formation;
\item $\alpha$, the 
average length of a new loop after it has formed.
\end{enumerate}
This last parameter is
under very active study at present from both analytical
 \cite{Polchinski:2006ee,Polchinski:2007rg, Dubath:2007wu}
and numerical \cite{Olum:2006ix,Vanchurin:2005pa,Martins:2005es,Ringeval:2005ce}
perspectives. It is an important thing to understand because large loops live much longer than 
small loops. Loops lose their length by generating gravitational radiation, which is now 
beginning to be brought under observational limits; since longer loops 
emit gravity waves later, they are more tightly constrained (see ref. \cite{Polchinski:2007qc}
for much more detail). For the discussion of our numerical results, we first
treat them as free and independent parameters so as to study
how each affects magnetogenesis. On the other hand, we take our test
values for $G\mu_0$ from observational constraints: a fiducial value of
 $2\times 10^{-7}$ \cite{Wyman:2005tu,Seljak:2006bg},
an optimistic value of $7\times10^{-7}$ \cite{Bevis:2007gh}, and
the most constrained value of $2 \times 10^{-8}$ \cite{Siemens:2006yp}; the former two
come from combining CMB data with other cosmological observations, while
the last is the ``worst case scenario" limit from pulsar timing.
At the end of this section, we present constraints on what we consider to be the
best motivated combination of parameters, which differs slightly from our fiducial model.
Our fiducial model, used wherever nothing else is specified, includes:
the VOS model for long strings; loop dynamics
turned on; $G\mu_0 = 2\times10^{-7}$; $\alpha=0.01$;
and $v_t(t=t_F) = 0.1$. Where relevant, we have drawn a line demarcating 
 the minimum correlation length necessary to seed galactic dynamos
and a line indicating the minimum magnetic field strength necessary, 
given a particular dynamic amplification time, $\Gamma_{dy}^{-1}$.

\begin{figure}[h!] 
   \centering
   \includegraphics[width=0.5\textwidth]{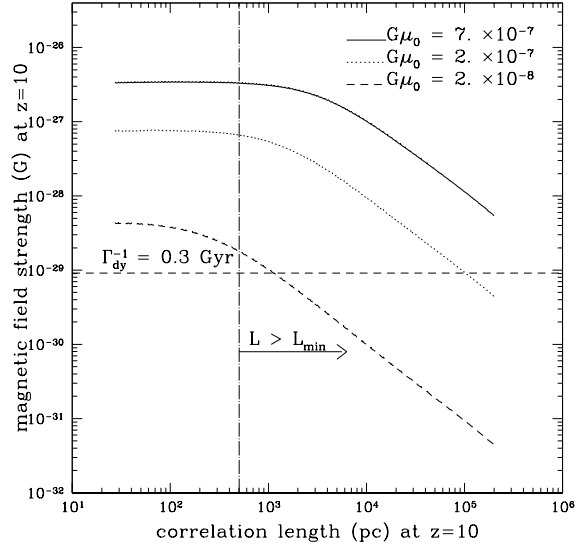} 
   \caption{Magnetic field strength as a function of the magnetic field's correlation length,
   for our three test values of bare string tension $G\mu_0$.  The plot shows
   the magnetic field strength at $z=10$ as a function of the proper correlation
   length at the same redshift.}
   \label{fig:Gmuvaried}
\end{figure}

\begin{figure}[hb!] 
   \centering
   \includegraphics[width=0.5\textwidth]{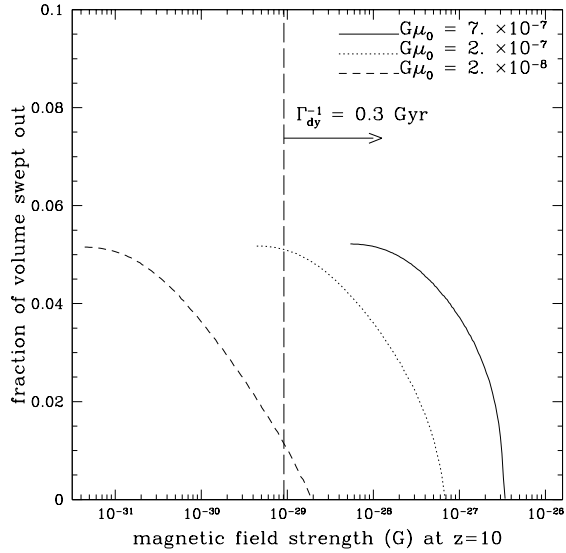} 
   \caption{Volume of universe suffused with a seed field as a function of seed field flux magnitude.
The horizontal axis gives the magnetic field strength  at $z=10$, and the
vertical axis shows the fraction of the volume of the universe which is
suffused with a magnetic field of that flux or greater.  
}
   \label{fig:vGmuvaried}
\end{figure}

\begin{figure}[h!] 
   \centering
   \includegraphics[width=0.5\textwidth]{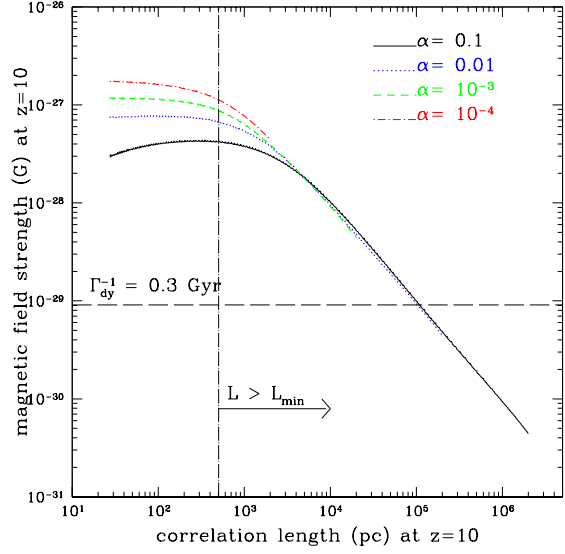} 
   \caption{Magnetic field strength as a function of the string formation length fraction $\alpha$. }
   \label{fig:alphavaried}
\end{figure}

\begin{figure}[h!] 
   \centering
   \includegraphics[width=0.5\textwidth]{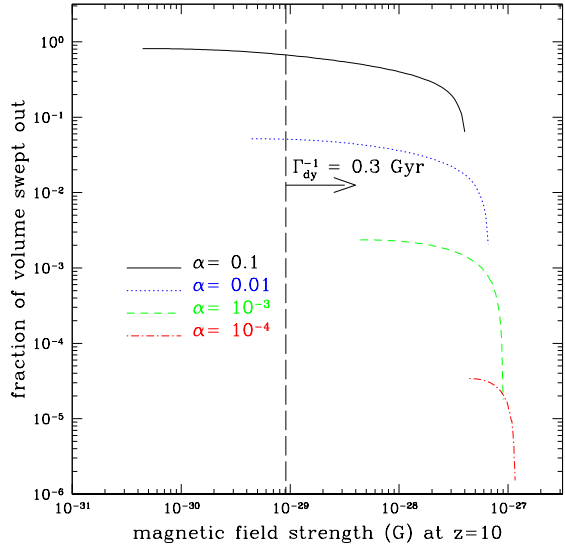} 
   \caption{Volume of universe suffused with a seed field as a function of seed field flux magnitude, for various values of $\alpha$.}
   \label{fig:valphavaried}
\end{figure}

\begin{figure}[h!] 
   \centering
   \includegraphics[width=0.5\textwidth]{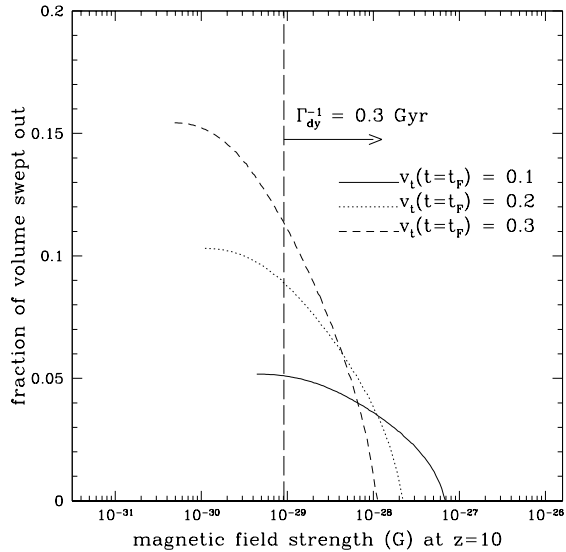} 
   \caption{Volume of universe suffused with a seed field as a function of seed field flux magnitude, for various values of the loop's initial translational velocity, $v_t(t=t_F)$.}
   \label{fig:varyut}
\end{figure}

\begin{figure}[h!] 
   \centering
   \includegraphics[width=0.5\textwidth]{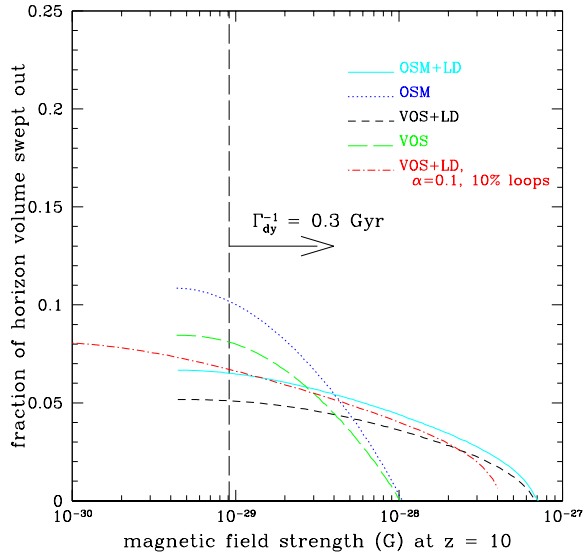} 
   \caption{ Since
   loops change their size over time, different volumes of space will be endowed with
   seed fields of differing magnitudes. Here we plot a line indicating the percent of the 
   universe covered by a magnetic field whose size is greater than or equal to some minimum 
   field strength. Results for both network models are shown for comparison, and for each we show
   results both with (+LD) and without loop dynamics. To explain a bit better, an example: the point where
   the red line crosses 8\% is at a magnetic field strength of approximately $10^{-30}$ G. What
   this means is that 8\% of the volume of the universe at the time of galaxy formation was suffused
   with a seed field whose magnitude was equal to, or larger than,  $10^{-30}$ G. 
   Here we have adopted $G\mu_0 = 2 \times10^{-7}$ and $\alpha = 10^{-2}$. Only fields whose
   correlation lengths are sufficiently large ($L_{corr} >500$pc at $z=10$) are included.  
   We have also included
   what we believe to be the best motivated model: the VOS model for the long
   strings, loop dynamics, and $\alpha=0.1$, but with only 10\% of the string network's 
   energy loss going into loops that large, with the rest lost to very small loops. }
   \label{fig:volumes}
\end{figure}

\begin{figure}[h!] 
   \centering
   \includegraphics[width=0.5\textwidth]{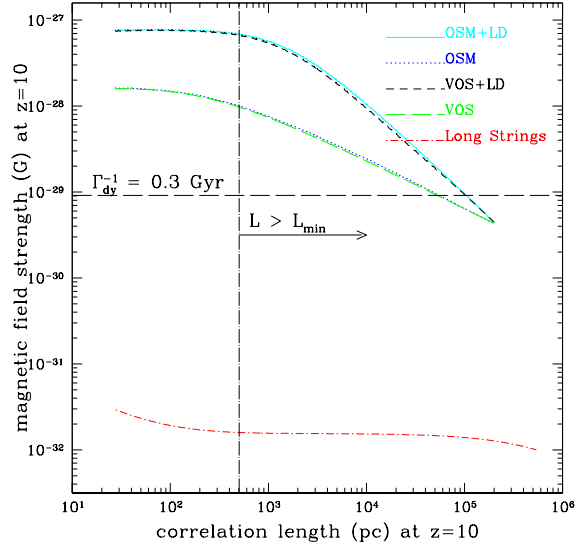} 
   \caption{Magnetic field strength for the different string models we consider, excluding our
   best motivated physical model (which on this plot is nearly identical with VOS+LD), compared
   with the prediction from long string encounters, where the vortex is created in the region
   between two oppositely moving long strings.}
   \label{fig:bfield}
\end{figure}

\begin{figure}[h!] 
   \centering
   \includegraphics[width=0.5\textwidth]{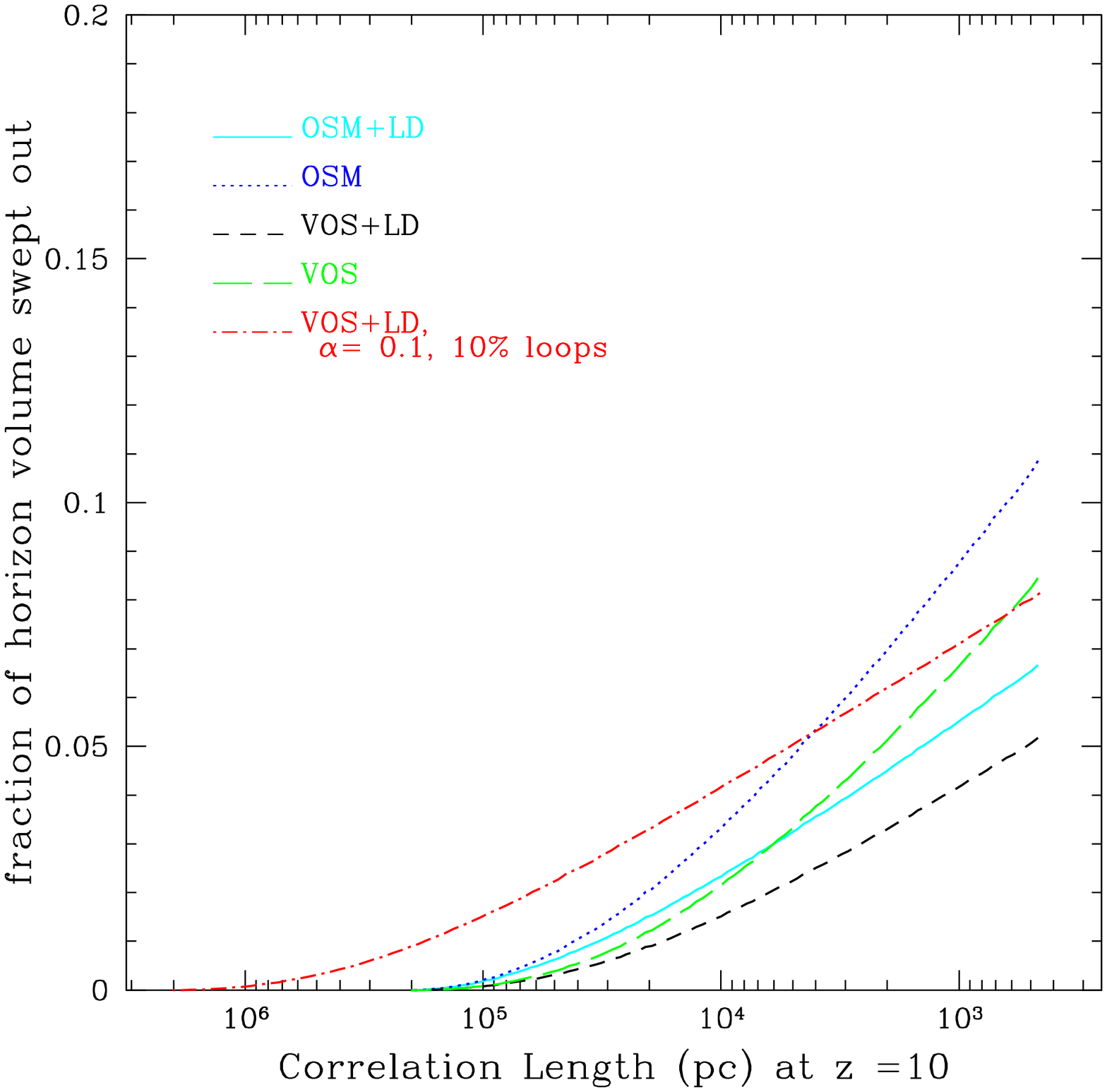} 
   \caption{For the five model combinations we consider, we plot the fraction of the volume
   covered versus the correlation length. Only fields whose
   correlation lengths are sufficiently large ($L_{corr} >500$pc at $z=10$) are included. }
   \label{fig:lenversusvolume}
\end{figure}

\subsection{Discussion}

We now discuss the results of the numerical simulations.  Unless stated otherwise, all of the plots assume the VOS model with loop dynamics, $G\mu_0 = 2\times10^{-7}$, $\alpha=0.01$,
and $v_t(t=t_F) = 0.1$. We have systematically explored the variation of our 
predictions with the parameters listed above. 

%
%

The scaling of the magnetic field spectrum with $G\mu_0$, illustrated in
Figures \ref{fig:Gmuvaried}, can be understood within the context of the
calculations in Sections \ref{s:Vorticity} and \ref{s:Networks}.  As 
$G\mu_0$ is reduced, each loop becomes less effective at generating magnetic
fields, and so the magnetic flux is reduced.  Furthermore, the
loop length spectrum peaks at a characteristic length determined by $G\mu_0$, as
given in (\ref{lpeak}) for the OSM model.  Since we use only a single
scale for loop formation here, we find the same result in the VOS model.
As $G\mu_0$ is reduced, this
characteristic loop size falls as well, shifting the magnetic field spectrum
to smaller correlation lengths.

The magnetic field strength as a function of correlation length for several
values of $\alpha$ is given in Figure \ref{fig:alphavaried}.  The magnetic
field spectrum is only weakly dependent on $\alpha$, thanks primarily to 
two effects.  First, while $\alpha$ sets the size of the largest loops, 
at any fixed time the greatest number of loops have characteristic length
that is set by $G\mu_0$, as described for the OSM model in (\ref{lpeak}).  Therefore the peak correlation length and magnetic field flux only weakly depend on $\alpha$.  Secondly, loops created in models with larger $\alpha$ will always shrink to smaller lengths, and thus mimic models in which $\alpha$ is smaller.  The peak magnetic flux grows slightly with $\alpha$, since at progressively smaller values of $\alpha$, the string network must shed energy into an ever greater number of ever smaller loops.  Thus, the fraction of loops creating fields at the peak correlation length grows slowly as $\alpha$ decreases.

To provide a viable magnetogenesis mechanism, cosmic string loops must not only produce sufficiently strong magnetic fields, but must produce them over nearly the entire volume of the universe.  Figures \ref{fig:vGmuvaried}, 
\ref{fig:valphavaried}, and \ref{fig:varyut} explore this issue for different values of 
$G\mu_0$, $\alpha$, and loop $v_t$ at formation.  In these figures, we have plotted the fraction of
a universe suffused with a given magnetic field strength or greater.  If
one chooses a value of magnetic flux required to seed a galactic dynamo, the curves give the fraction of the universe in which the string network produces the required seed fields.  Figure \ref{fig:vGmuvaried} shows the same fall in
 peak magnetic field strength with falling $G\mu_0$ as does Figure \ref{fig:Gmuvaried}.  Beyond shifting to smaller values of magnetic field strength, the curves look roughly similar: as $G\mu_0$ is decreased, the loop network 
 contains roughly the same number and sizes of loops, which produce the same 
volume coverage but lower magnetic field strengths.  Figure
\ref{fig:valphavaried} shows the same weak dependence of the peak magnetic
field strength on $\alpha$ as does Figure \ref{fig:alphavaried}.  The volume 
coverage falls nearly linearly with $\alpha$, as is to be expected since at 
smaller values of $\alpha$ the loop network contains fewer large loops.
The scaling, very roughly speaking, is that the volume swept out by a single loop
is proportional to $\alpha^2$, while the number of loops produced is proportional
to $\alpha^{-1}$, the combination of which gives an approximately linear scaling
of volume with $\alpha$. Finally, in \ref{fig:varyut}, we see an approximately
linear scaling in volume coverage with loop initial transverse velocity. Faster loops
can sweep out more volume, but since they remain in any one area for a shorter 
period of time, they produce weaker magnetic fields. This is also evident in the plot.

Figures \ref{fig:volumes}, \ref{fig:bfield}, and \ref{fig:lenversusvolume}
 explores the effect of loop dynamics and the network model on our conclusions. 
In Figure  \ref{fig:volumes}, we plot the horizon volume overswept
versus the magnetic field strength, as in Figures \ref{fig:vGmuvaried}, \ref{fig:valphavaried}, 
\ref{fig:varyut} above. Two effects are evident in these results. The One Scale Model
covers a slightly large volume of the universe with fields than the Velocity-dependent
One Scale model, as the OSM tends to overestimate the number of long strings present
during the matter-radiation transition era, leading to a greater number of loops
present during magnetogenesis. The inclusion of loop dynamics increases 
magnetic field strength but decreases volume coverage, since loop dynamics
decrease loop translational velocity over time. In addition to these model
comparisons, we also plot the model we believe to have the best physical motivation:
the VOS model, with loop dynamics included, where string loops are formed
at 10\% of the horizon size, but with only 10\% of the loop energy going into 
loops this large. We assume that the remaining 90\% of loops are formed near the
gravitational radiation back-reaction scale, through loop fragmentation or other
effects \cite{Polchinski:2007qc}. In Figure \ref{fig:bfield}, we compare
the magnetic field strengths in our models with the magnetic fields expected
from long string encounters. Since the vortices formed by long
string encounters are spread over such large length scales, the fields
generated by those encounters are concomitantly much weaker
than those generated by loops. Finally, in Figure \ref{fig:lenversusvolume},
we plot volume overswept versus correlation length for these same five models,
beginning with the largest correlation lengths.
This plot includes only field correlations lengths that are sufficiently large ($L_{corr} > 500$pc
at $z=10$).

\section{Conclusions}\label{s:Conclusions}

In this paper, we have studied the viability of a cosmic string network for generating the seed
fields necessary for explaining, after dynamic amplification,
 the $\mu G$ magnetic fields observed in spiral galaxies. 
We have analyzed the behavior of wiggly cosmic strings in the era between matter-radiation
equality and decoupling. In this era, we find that strings can create vorticity
in the primordial plasma, allowing the Harrison-Rees mechanism for magnetogenesis
to operate. This effect is, however, at second order in the string's effective Newtonian
mass density, contrary to earlier estimates in the literature. Though the turbulent eddies in
wakes behind long strings can form stronger magnetic fields, by our calculations
these fields are on length scales too small to explain galactic magnetic fields. On the other
hand, we find only weak support in our calculations for vortices forming in the space between
distantly separated straight strings, which would have adequate correlation lengths. 

However, by interpolating between these extremes, we have discovered that cosmic string loops 
provide an excellent mechanism for magnetogenesis, with correlation
length scales that are large enough as well as field strengths much greater
than those predicted for the speculative inter-string vortices.  If we take present
work on loop production seriously, then approximately 10\% of the energy a string network
loses goes into strings whose length is approximately $0.1 L_H$. These loops can oversweep
approximately 10\% of the universe during the relevant epoch between matter-radiation equality
and decoupling. With loops preferentially oversweeping areas with overdensities,
this may be just enough coverage. Also, if cosmic strings are $(p,q)$ strings from 
string theory, we can expect a further enhancement of this volume coverage by a
factor of approximately $3 P^{\beta}$ \cite{Tye:2005fn}.
In those regions that are overswept, 
strings with tensions allowed by present-day cosmological constraints can
generate seed magnetic fields large enough to account for today's 
magnetic fields under the assumption of an efficient galactic dynamo for amplification.
Although we are making optimistic assumptions about these dynamos, we have consistently
made conservative estimates for the string-driven magnetogenesis.
For instance, we are not forced to rely on turbulence or inverse cascades to achieve these results;
if such mechanisms exist, then the seed fields we predict would only be strengthened.

We achieved these predictions first through careful analytical estimation, then through the
construction of a thorough computer code. This latter has allowed us to simultaneously
solve for the evolution of the string network, loop production, loop dynamics, and cosmology,
giving us predictions for magnetic fields that take into account a multitude of effects. Beyond
a bare estimate of the magnetic field amplitude, we correlate the volume
of the universe overswept by loops with such parameters as $G\mu_0$, the string tension,
and $\alpha L_H$, the typical size of loops. In addition, we compare the results of two
popular semi-analytic string network models, the original one scale model
as well as the more sophisticated velocity-dependent one scale model. Because
of the versatility supplied by the code, we predict, inter alia,
the spectrum of magnetic field strength versus correlation length.
Though it may be far fetched,
if future astronomical observations were capable of detecting and characterizing the primordial 
seed field, our predictions are sufficiently concrete that such observations
could either confirm or refute our model.

In addition to futuristic observations, our model could also be refuted through the definitive
disproof of efficient galactic dynamos (our model requires $\Gamma_{dy}^{-1} \lesssim 0.3$),
or the constraint of cosmic string tensions to much less than $G\mu \simeq 10^{-8}$. 
On the other hand, if cosmic strings are observed and possess a tension of around our
fiducial value, the case for this magnetogenesis mechanism 
would be greatly strengthened. 

Thus, the presence of a string network might help explaining some large scale magnetic fields if efficient galactic dynamos are present.
 Put in other words, with efficient galactic dynamos, 
we can explain a $\mu \text{G}$ with a 
$G\mu$.

\begin{acknowledgments}
We thank S.~Alexander for early discussions motivating this project as well as A.~C.~Davis and K.~Dimopoulos for feedback, especially regarding the role of turbulence in string wakes. 
D.W. and M.W. thank the University of Sussex for their wine and hospitality
during the completion of this work.
 T.~B. is supported by PPARC grant PP/D507366/1 and thanks the Perimeter Institute for their hospitality.
The work of M.~W. at the Perimeter Institute is supported in part by the Government of Canada
through NSERC and by the Province of Ontario through MEDT. D.~B. would like to thank A.~C.~Davis and DAMTP for support.
\end{acknowledgments}

\appendix
\section{Dynamics of rotating loops \label{ap1}}

Loops are acted upon by 
dynamical friction forces as well as the recoil and 
shrinking from the emission of gravitational
waves.  We must estimate these effects to ensure 
that the approximations we make in Section \ref{sec:rotloop} are valid,
as well as to properly incorporate them in the numerical estimates
presented in Section \ref{ss:NumericalResults} and described in 
detail in Appendix \ref{ap2:code}.
In this Appendix we derive the differential equations that govern
the evolution of the 
translational velocity $v_t(t)$, rotational velocity $v_r(t)$ and
 length $\ell(t)$ of a loop, and provide approximate analytic solutions.  

\subsection{Changes in size and shape \label{ap1:shrink}}
The length of the loop decreases due to gravitational radiation,
so that 
\begin{eqnarray}
\ell(t)&=&f_r\alpha L_H(t_F)-\Gamma_l G\mu_0(t-t_F)\\
&\equiv&\ell_0-G\Gamma_l\mu_0(t-t_F)\label{loft}\,,
\end{eqnarray}
where $f_r\leq 1$ describes energy loss directly after
formation, $\Gamma_l\approx 50$ controls the efficiency with which the
loop emits gravitational radiation \cite{CaldwellAllen1992,Allen:1994iq,Casper:1995ub},  $\ell_0\equiv\tilde{\alpha}t_F=f_r\alpha L_H(t_F)$, and  
 the time scale for loop shrinkage is
\begin{eqnarray}
t_{shrink}\equiv \frac{\tilde{\alpha}}{G\Gamma_l\mu_0}t_F= 2000 t_F \,,  \label{tshrink} 
\end{eqnarray}
for $\tilde{\alpha}=0.01$ and $G\mu_0=10^{-7}$; so a loop
with $\tilde\alpha > 10^{-5}$ remains large enough for magnetogenesis for 
many Hubble times.

A further concern is large-scale loop oscillations, which 
change the shape of the loop significantly.
(Small scale oscillations of the string are already averaged over to give
the effective tension $T$ and linear mass density $\mu$). 
If the oscillation timescale is comparable
to the rotation period, the rigid loop approximation does not apply. 
Since the initial velocity distribution on the loop has to be quite peculiar to yield a fast oscillation affecting the loop as a whole, we expect that 
these oscillations make
only a small fraction of loops unsuitable.

\subsection{Translational Movement \label{ap1:trans}}

Three effects determine the drift velocity of a loop: redshifting from Hubble expansion, dynamical friction due to dragging of the plasma as computed in (\ref{accdrag}), and recoil from gravitational wave emission
 \cite{Vachaspati:1984gt} (often referred to as the {\it rocket effect}). 
The latter causes an acceleration of $\Gamma_p G \mu_0 \hat{{\bf n}}/\ell$, where $\hat{\bf{n}}$ is a unit vector in the direction of recoil and $\Gamma_p\approx 10$ \cite{Vachaspati:1984gt}. 
(It is $\mu_0$ and not $\lambda=\mu-T=\mu(1-\mu_0^2/\mu^2)\approx 0.56\mu_0$ which determines the emission of gravitational waves.)
Incorporating these forces
leads to
\begin{eqnarray}\label{eq:LoopTranslationalVelocity}
\dot{{\bf v}}_t=-H{\bf v}_t-\frac{{\bf v}_t \ln \theta_{min}^{-1}}{t_*}+\frac{\Gamma_p G \mu_0}{\ell}\hat{{\bf n}}, \label{Gammap}
\end{eqnarray}
where 
$t_*$ is defined in (\ref{eq:TStar}) and $\theta_{min}$ in (\ref{thetamin}). 
In the matter era, using $H=2/(3t)$ and $8\pi G \rho=3H^2$
gives
\begin{eqnarray}
t_*\equiv \frac{v_t^3t^2}{C_1}\,, \quad
C_1 = \frac{2}{3}G\ell \lambda\,.
\end{eqnarray}
Since our arguments in the previous subsection show that 
the loop length $\ell$  decreases very little over the 
timescales we are interested in, we take $\ell$ to be a constant.  Further, we ingore the time dependence in the logarithm and estimate this factor by \footnote{We approximate $r_{max}=\int v_t\,\text{d}t\approx 3v_t t\approx 3v_t(t_F)t_F$, anticipating $v_t\propto 1/a\propto 1/t^{2/3}$ initially, due to redshifting.}
\begin{eqnarray}
\ln \theta_{min}^{-1}&\approx&\ln\left(\frac{3v_t^3(t_F)t_F}{2G\lambda\ell}\right)=\text{const}\,,\label{approxln}
\end{eqnarray}
where the time of loop creation is 
\begin{eqnarray}
t_F=\ell/\tilde{\alpha}.
\end{eqnarray} 
Different initial values can lead to
great differences in the
 long-term behavior of
$v_t$. 
The loop either slows rapidly, or accelerates quickly
to relativistic velocities from the rocket effect.
Below we 
integrate the dynamical equation (\ref{eq:LoopTranslationalVelocity}) 
in each regime, and derive the
limiting velocity $v_{lim}$ that separates the two regimes.
Without the rocket effect term, equation (\ref{eq:LoopTranslationalVelocity})
becomes
\begin{eqnarray}
\dot{v}_t=-\frac{2}{3t}v_t-\frac{C_1\ln(\theta_{min}^{-1})}{t^2v_t^2}\,,
\end{eqnarray}
which has the solution
\begin{eqnarray}
\nonumber v_t^{f}(t)&=&t^{-2/3}\left(-3C_1\ln(\theta_{min}^{-1})t+\frac{\ell}{\tilde{\alpha}^2}(3C_1\ln(\theta_{min}^{-1})\tilde{\alpha}+v_0^3\ell)\right)^{1/3}\\
&\approx& v_0\left(\frac{\ell}{\tilde{\alpha} t}\right)^{2/3}\left(1-\frac{t}{t_f}\right)^{1/3}
\end{eqnarray}
where the last step uses  $v(t_F)=v_0$, neglects $3C_1\tilde{\alpha}\ln(\theta_{min}^{-1})$, and employs the time scale of dynamical friction  defined by
\begin{eqnarray}
t_f\equiv\frac{v_0^3\ell}{2\lambda G\tilde{\alpha}^2\ln(\theta_{min}^{-1})}\,.
\end{eqnarray}
When  dynamical friction is irrelevant 
(\ref{eq:LoopTranslationalVelocity}) becomes
\begin{eqnarray}
\dot{v}_t=-\frac{2}{3t}v_t+C_2\,,
\end{eqnarray}
where $C_2\equiv\Gamma_pG \mu_0/\ell$ and we assume that the recoil is  collinear with the velocity.
This has the solution
\begin{eqnarray}
\nonumber v_t^{r}(t)&=&\frac{1}{5t^{2/3}}\left(3C_2t^{5/3}-\frac{\ell^{2/3}}{\tilde{\alpha}^{5/3}}\left(3\ell C_2-5v_0\tilde{\alpha}\right)\right)\\
&\approx&v_0\left(\frac{\ell}{\tilde{\alpha} t}\right)^{2/3}\left(1+\left(\frac{t}{t_r}\right)^{5/3}\right)\,,
\end{eqnarray}
where we neglect $3\ell C_2$ in the last step and introduce the relevant time scale $t_r$ for the rocket effect  
\begin{eqnarray}
t_r\equiv\left(\frac{5v_0}{3\Gamma_p G\mu_0}\right)^{3/5}\frac{\ell}{\tilde{\alpha}^{2/5}}\,.
\end{eqnarray}
Setting $t_r=t_f$ and solving for the initial velocity yields 
\begin{eqnarray}
v_{lim}=G^{1/6}\left(\frac{\lambda^5}{\mu_0^3}\right)^{1/12}\left(\frac{5}{3\Gamma_p}\right)^{1/4}2^{5/12}\tilde{\alpha}^{2/3}(\ln \theta_{min}^{-1})^{5/12}\,,\label{vlim}
\end{eqnarray}
Hence, for $v_t\ll v_{lim}$ dynamical friction is more important, whereas
for larger velocities, the rocket effect predominates in the
long run.

What is a reasonable initial value for $v_t$?
If large loops are created in the matter era, we expect their
 velocities to be  comparable to the RMS velocity in the string network
$v_{RMS} \sim 0.6$ \footnote{Recent simulations and analytic arguments suggest two distinct classes of loops \cite{Polchinski:2007rg,Polchinski:2007qc}: small, highly relativistic loops, and large ones, which have the velocities we consider.}. If we  average over the small scale wiggles, the RMS velocity drops down to ${\bar v}_{RMS} \sim 0.15$ \cite{Vachaspati:1991tt}, so the initial translational velocity of the loops should be smaller too.
Rotating loops share kinetic energy between translation and rotation, 
so an initial value of $v_t\sim v_r$, say down to $0.4$, may
be reasonable. This velocity is still larger than the limit velocity 
(\ref{vlim}), even for the largest loops produced in the network: 
for $\tilde{\alpha}=0.01$ and the bound $G\lambda/0.56=G\mu_0\lesssim10^{-7}$, the limiting velocity becomes $v_{lim}\lesssim 0.0072$, where we approximated $\ln \theta_{min}^{-1}\approx 19$, based on (\ref{approxln}) with $v_t(t_F)=0.4$.
Hence, the translational velocity of loops which are
created in the matter era  is given by
\begin{eqnarray}
v_t(t)&\approx&v_0\left(\frac{\ell}{\tilde{\alpha} t}\right)^{2/3}\left(1+\left(\frac{t}{t_r}\right)^{5/3}\right)\,,
\end{eqnarray}
for all relevant values of $\tilde{\alpha}$. The loop
initially slows due to redshifting, even though the rocket
effect dominates over dynamical friction. Indeed, for $\tilde{\alpha}\sim
0.01$, $v_0=0.4$ and $\Gamma\mu_0=10^{-7}$ we have $t_r\approx 197 \ell/\tilde{\alpha}= 197\, t_F$.
In
this case, $v_t$ drops down to the minimal value of  $v_t\approx 0.024$ at $t\approx t_r 3/4 $, and it increases linearly thereafter up until loop shrinking becomes important. Since it takes more than $1000\,t_F$ for a loop to accelerate so that it moves faster than $v_0$ again, we expect 
the majority
of loops to have a translational velocities of order $v_t\sim\mathcal{O}(10^{-1})$, which we
 use as a rough estimate of $v_t$ for loops created between $t_{eq}$ and $t_{dec}$. Since $t_r<t_{shrink}$ from (\ref{tshrink}), 
 our assumption of a fixed loop length $\ell$
is justified.

\subsection{Rotational Movement \label{ap1:rot}}

The rotational velocity $v_r$ of a loop is influenced by three effects:
dynamical friction from plasma drag,
emission of gravitational radiation \footnote{There is no rocket effect for angular momentum: numerical studies show that the emission of gravitational waves always decreases the angular momentum \cite{Durrer:1989zi}, even though a rigorous mathematical proof is lacking. The fundamental mass density $\mu_0$ determines this effect and not the effective mass density $\lambda$. } 
which produces a torque \cite{Durrer:1989zi}
\begin{eqnarray}
\tau_{gr}=-\ell G\mu_0^2\Gamma_{gr}\, \label{Gammagr}
\end{eqnarray}
where $\Gamma_{gr}\approx 5$ \cite{Durrer:1989zi}, and the
shrinking of the loop from gravitational wave energy emission.
The torque $\tau_{drag}$ from dynamical friction
is computed using the arguments of Section \ref{sec:rotloop}, applied 
to a single loop rotation. During one rotation, spanning a
time $\Delta t \sim \ell/v_r$,
the surrounding plasma
acquires an angular momentum of roughly

\begin{eqnarray}
\Delta J\sim \ell\Delta p_{plasma}\sim 
\ell^4\rho v_x\sim \frac{\ell^4 \rho v_y^2}{7 v_r} 
\end{eqnarray}
which gives the torque
\begin{eqnarray}
\tau_{drag}\approx - \frac{(2\pi)^2}{7} \frac{G^2\lambda^2}{v_r^2}\ell^3\rho .
\end{eqnarray}
Loop shrinkage enters via the expression for the 
total angular momentum $J$,
\begin{eqnarray}\label{eq:LoopRotationalVelocity}
\dot{J}=\frac{\lambda}{4\pi}\left(2\ell\dot{\ell}v_r+\ell^2\dot{v}_r\right)\,,
\end{eqnarray}
with $\ell(t)$ from (\ref{loft}). We neglect redshifting of the rotational 
velocity for loops well within the horizon.

Unlike the translational velocity, it is possible to obtain analytic
solutions for the rotational velocity with dynamical $\ell(t)$.
Dynamical friction is negligible, since the ratio of torques is
\begin{eqnarray}
\frac{\tau_{drag}}{\tau_{gr}}=\frac{(2\pi)^2}{7} \frac{\lambda^2 \tilde{\alpha}^2}{\mu_0^2v_r^2\Gamma_{gr} 6\pi}\frac{t_F^2}{t^2} \approx 1.3\times 10^{-5}\frac{t_F^2}{ t^2}\ll 1
\end{eqnarray}
where we use $8\pi G\rho=4/(3t^2)$ as well as $\lambda/\mu_0\sim 0.6$, $\tilde{\alpha}\sim 0.01$, $v_r\sim 0.4$ and $\Gamma_{gr}\approx 5$.
This leaves $\dot{J}=\tau_{gr}$, leading to
\begin{eqnarray}
\dot{v}_r=\frac{2 G\Gamma_l\mu_0 v_r(t)-C_{gr}}{\ell_0-G\Gamma_l\mu_0(t-t_F)}\,,
\end{eqnarray}
where
\begin{eqnarray}
C_{gr}=\frac{4\pi G\mu_0^2\Gamma_{gr}}{\lambda}\,,
\end{eqnarray}
and $\ell(t)$ is given by (\ref{loft}).
Assuming the initial condition 
$v_r(t_F)=v_0$, then $v_r(t)$ is
\begin{eqnarray}
v_r(t)=\frac{1}{2}\frac{C_{gr} f(t)+2v_0\tilde{\alpha}^2}{G\Gamma_l\mu_0f(t)+\tilde{\alpha}^2}
\end{eqnarray}
with
\begin{eqnarray}
f(t)\equiv 2\tilde{\alpha}\left(1-\frac{t}{t_F}\right)+G\Gamma_l\mu_0\left(1-\frac{t}{t_F}\right)^2\,.
\end{eqnarray}

As with the translational velocity $v_t$, there is a critical initial
rotational velocity that determines the future evolution of $v_r$, given by
$v_{lim}^r\equiv C_{gr}/(2G\Gamma_l\mu_0)$. For the parameters in table \ref{table:Parameters} we get $v_{lim}^r>1$, indicating that $v_r$ decreases for all loops under consideration. 

What is the initial velocity of a loop?
Loops can be produced by self-intersections of a single
string. In this case, we expect the intersecting pieces to move roughly in the same direction. As a consequence, the majority of the energy will go into translational movement and a loop
with very little angular momentum but large momentum results, so that $v_r \ll v_t$. 
However, Loops can also be created when two
strings, moving in opposite directions, intersect
 and chop off a loop. Loops formed in this way have very little momentum, but large anglar momentum, so that initial
rotational velocities comparable to the RMS velocity of the network result
$v_r\approx v_{RMS}$, while $v_t \ll v_r$. 

An ideal loop for magnetogenesis lies
somewhere in between. 
Optimally efficient loop magnetogenesis requires  
 rotation to stir up the
plasma, but also rapid translational motion  to oversweep a large
fraction of the universe. 
To account for kinetic energy in the form of the small scale wiggles and oscillations of the loop, as well as the angular momentum that is radiated away in
gravitational waves immediately after its creation, we take $v_r\sim 0.4<v_{RMS}$ to be a conservative initial value.

\section{Code Implementation \label{ap2:code}}

To fully account for the background cosmological model, the forces and torques 
on rotating loops, and the behavior of the string/loop network when the universe
is not in a scaling regime, we employ a computer program which implements the 
full set of differential equations described in this paper.  
Here we discuss the operation of this code.

The code follows an array of loop cohorts over time,
 while simultaneously  tracking some averaged quantites describing the
cosmological model and network of long strings.  At each time step, the code 
updates the background cosmological quantities (Hubble parameter, energy 
density components, \emph{etc.}) according to the Friedmann and 
energy conservation 
equations.  It also updates the properties of the long string network by using
either the one-scale model (Section \ref{ss:OSM}) or the velocity-dependent 
one-scale model (Section \ref{ss:VOS}).  Both the OSM and VOS models include 
a production rate for loops at a size determined by the background cosmological
parameters, so in each time step the appropriate number of loops
 are added to the corresponding cohort in the array.

In each time step 
we include all of the physics of the loop dynamics and magnetic field 
generation for each loop cohort.  
Each individual
cohort collects all the loops created during a block of time steps.  We
typically use  $10^6$ steps equally spaced in $\log(t)$ from 
redshifts $z \approx 10^8$ until the present, with
blocks consisting of $10^3$ steps each.  The loops
created during each time block have similar properties, but there are
slight differences between the loops added at the beginning and end of
a single block. The code
accounts for these differences by updating some of the  cohort
 parameters 
 through a weighted average of the properties of the newly added loops and the 
properties of the  loops already in the cohort.  Each cohort tracks a number of different variables:

\begin{enumerate}

\item The comoving number density of loops in the cohort,
 as computed by the OSM (\ref{eq:OSMLoopProdRate}) or VOS equations
(\ref{eq:VOSLoopProdRate}).  The comoving number density is  not
averaged at each time step, but accumulates new loops as they are created.

\item The physical length of the loops in the cohort,  which
decreases due to the emission of gravitational radiation, with the rate
given by (\ref{eq:LoopLenFunTime}).  This is averaged over the cohort.

\item The translational and rotational velocity of the loops, which  are 
computed using (\ref{eq:LoopTranslationalVelocity}) and (\ref{eq:LoopRotationalVelocity}), respectively.  This includes Hubble
damping, the rocket
effect, dynamical friction, torque from graviatational wave emission, and 
the shrinking moment of inertia of the loops.  These properties are 
averaged over the
cohort.

\item The fractional comoving volume overswept by the loops in the cohort, which
depends on the comoving number density and translational velocity of the loops.
This is  a proxy for the fraction of the universe over which
the loops in the cohort may generate magnetic fields, and is accumulated (not
averaged) as new loops are added.  

\end{enumerate}
At each time step, the program updates each of these quantities, from the 
time the first loops are added to the cohort until the cohort has 
completely evaporated
into gravitational radiation.

The ultimate objective is  to calculate the magnetic field strength
as a function of scale created by the loops.  This is tabulated 
in an array which tracks the  
comoving magnetic fluxes $a^2(t)B(t)$ on a 
selection of length scales.
Each magnetic field bin
also tracks the fractional comoving volume of the universe which contains 
magnetic fields at that length scale. 
An individual
 loop cohort generates magnetic fields on large comoving
scales when the loops are newly created, and then on progressively smaller 
scales as the loops shrink.  
Furthermore, the magnetic field at each comoving
correlation length gets contributions from many different loop cohorts over
time.  
  Therefore the array tracking magnetic field strengths
is indexed differently from the one tracking the loop cohorts,
with each array 
element corresponding to a fixed range of comoving lengths.  

At each time step, the code steps through the loop cohorts, 
and assigns the magnetic fields 
generated by each cohort in that time step to a bin in the magnetic 
field array, 
determined by the size of the loops in the cohort.   The comoving magnetic
flux is then updated by averaging the flux already present and the flux
created by the loops in the cohort, weighted by the comoving volume traversed
by the loops in the cohort and the total comoving volume already present in
the bin.  
The comoving volume represented by the bin is incremented by the volume
traversed by the loop cohort over the last time step.
At the end of the simulation each bin in the magnetic field array 
represents the average magnetic field strength in a certain fraction of the 
universe over which the field is correlated over a given length.
 
To compare the field created by the loop population to that created by
the long strings, a parallel calculation tracks the magnetic field created
by the long string population.  This is much simpler than the loop 
calculation, primarily because the 
correlation length for long-string magnetic fields is given by the interstring
separation, which is usually some nearly constant fraction of the horizon
length.  Thus each comoving correlation length only gets contributions from 
a fixed moment in cosmic history.


\begin{center}
\begin{table}[tb]
\begin{tabular}{llll}
\hline\hline
Parameter&Value& Source& description, first used in (equation)\\
\hline
$G\mu_0$ &$ 2\times 10^{-7}$ & \cite{Wyman:2005tu,Seljak:2006bg}& string mass/length \\
$\mu$ & $1.9\mu_0$ & \cite{VilenkinShellard2000}, radiation era& effective mass/length (\ref{muzeroandmu})\\
 & $1.5\mu_0$ & \cite{VilenkinShellard2000}, matter era& \\
$\alpha$&$0.01$ & \cite{CaldwellAllen1992,VilenkinShellard2000}& size of large loop/horizon (\ref{eq:LoopLenFunTime})\\
 $f_r$  &$0.7$ & \cite{BennettBouchet,CaldwellAllen1992}& loop redshifting
 energy loss (\ref{eq:LoopLenFunTime})\\
$\tilde{\alpha}$&$f_r\alpha H^{-1}(t_F)/t_F$ & Defined in (\ref{eq:TildeAlphaDef}) & new loop length over formation time\\
$\Gamma_{\ell}$& $50 \quad (\star)$& 
& efficiency of grav. wave emission \cite{Hindmarsh:1994re} (\ref{eq:LoopLenFunTime})  \\
& $50\lesssim \Gamma_{\ell}\lesssim 100$& \cite{Allen:1994iq}
&  \\
&$45\lesssim \Gamma_{\ell}\lesssim 55$& \cite{Casper:1995ub}
 &  \\
 & $50$, $80$& \cite{CaldwellAllen1992,Durrer:1989zi,Hogan:1984is}
&\\
$\Gamma_{gr}$ & $5$ & \cite{Durrer:1989zi}& grav. wave emission yielding torque (\ref{Gammagr})  \\
$\Gamma_{p}$ & $10$ &\cite{Allen:1994ev,Vachaspati:1984gt,Durrer:1989zi} &rocket effect (\ref{Gammap})   \\
$\Gamma_{dy}$ & 0.2 $\text{Gyr}\quad(\star)$ & 
& dynamo efficiency (\ref{eq:Gammady})\\
 & $0.2<\Gamma_{dy}^{-1}/\mbox{Gyr}<0.8$ & 
\cite{Ruzmaikin, Davis:1999bt}& \\
& $\Gamma_{dy}^{-1}\gtrsim (1.1-1.4)\,\mbox{Gyr}$ & \cite{Widrow:2002ud}& \\
& $\Gamma_{dy}^{-1}=0.3\,\mbox{Gyr}$ &  \cite{RST}& \\
& $\Gamma_{dy}^{-1}=0.5\,\mbox{Gyr}$ &  \cite{dynamo}& \\
& $\Gamma_{dy}^{-1}=2.2\,\mbox{Gyr}$ &  \cite{Ferriere:2000}& \\
$z_{dec}$ &  $1089$ & \cite{Spergel:2006hy}& redshift at decoupling \\
$z_{gf}$ &  6 & \cite{Dimopoulos:1997df}& redshift of galaxy formation\\
         & 10 $(\star)$ & & \\
$t_{eq}$ & $1.6\times 10^{12}$ s = 51 kyr & & time of matter-radiation equality 
  \\
$t_{dec}$ & $1.2\times 10^{13}$ s = 380 kyr & &time of decoupling (\ref{finalBs})\\
$t_{0}$ & $ 13.7\, \mbox{Gyr}$ &\cite{Spergel:2006hy}& age of the universe \\
$v_{RMS}$ & $0.60$ & \cite{Vachaspati:1991tt}, matter era& RMS velocity in network  
\\
$\bar{v}_{RMS}$ & $0.15$ & \cite{Vachaspati:1991tt}, matter era&  RMS velocity avg. over corr. length \\
$\Omega_m h^2$ & $0.1277$ & \cite{Spergel:2006hy}& matter fraction (\ref{xcorr}) \\
$h$ & $0.732$ & \cite{Spergel:2006hy}& Hubble parameter (\ref{xcorr})\\
$\beta$ & $1/2\leq\beta\leq 1$ & \cite{Davis:2005ih}& (\ref{mentionP}) \\
$P$ & $1$ $(\star)$& cosmic strings& intercommutation probability \\
 & $10^{-1}\leq P \leq 1$ & D-Strings \cite{Polchinski:2004ia}& \\
 & $10^{-3}\leq P \leq 1$ & F-Strings \cite{Polchinski:2004ia}& \\
 $c_1$ & 0.21 (0.2475) & \cite{Martins, Tye:2005fn} & VOS parameters in radiation (matter) era\\
 $c_2$ & 0.18 (0.3675) & & \\
 $c_3$ & 0.28 && \\
 
\hline\hline
\end{tabular}
\caption{Our parameters.  When several
values are given, we select the one marked by
a $(\star)$.}
\label{table:Parameters}
\end{table}
\end{center}


\end{document}